\documentstyle[12pt]{article}

\input psfig.sty

\renewcommand{\baselinestretch}{1.2}
\newcommand{\nc}{\newcommand}

\nc{\eqr}[1]{(\ref{#1})}
\nc{\sref}[1]{\S\ref{#1}}
\nc{\tref}[1]{Table~\ref{#1}}
\nc{\fref}[1]{Figure~\ref{#1}}
\nc{\cref}[1]{Chapter~\ref{#1}}
\nc{\beq}{\begin{equation}}
\nc{\eeq}{\end{equation}}
\nc{\barray}{\begin{eqnarray}}
\nc{\earray}{\end{eqnarray}}
\nc{\barrayn}{\begin{eqnarray*}}
\nc{\earrayn}{\end{eqnarray*}}
\nc{\bcenter}{\begin{center}}
\nc{\ecenter}{\end{center}}
\nc{\lra}{\longrightarrow}
\nc{\ra}{\rightarrow}
\nc{\setall}{\setcounter{equation}{0}
        \setcounter{definition}{0}
        \setcounter{lemma}{0}
        \setcounter{convention}{0}
        \setcounter{conjecture}{0}
        \setcounter{theorem}{0}
        \setcounter{proposition}{0}
        \setcounter{property}{0}
        \setcounter{fact}{0}
        \setcounter{corollary}{0}}
\nc{\setequation}{\setcounter{equation}{0}}
\nc{\hs}[1]{\hspace{#1 mm}}
\nc{\PSbox}[3]{\mbox{\rule{0in}{#3}\includegraphics{#1}\hspace{#2}}}

\def\sCC{{\kern 0.27em\vrule height1.45ex width0.03em depth0em
          \kern-0.30em\rm C}}
\def\C{{\mathchoice
  {\sCC}
  {\sCC}
  {\kern 0.225em \vrule height1.05ex width0.025em depth0em \kern-0.25em \rm C}
  {\kern 0.180em \vrule height0.78ex width0.02em depth0em \kern-0.2em \rm C}
        }}
\def\sHH{{\rm I\kern-.16em{}H}}
\def\H{{\mathchoice
  {\sHH}
  {\sHH}
  {\rm I\kern-.13em{}H}
  {\rm I\kern-.13em{}H} }}
\def\sNN{{\rm I\kern-.16em{}N}}
\def\N{{\mathchoice
  {\sNN}
  {\sNN}
  {\rm I\kern-.12em{}N}
  {\rm I\kern-.10em{}N} }}
\def\sPP{{\rm I\kern-.16em{}P}}
\def\P{{\mathchoice
  {\sPP}
  {\sPP}
  {\rm I\kern-.12em{}P}
  {\rm I\kern-.10em{}P} }}
\def\sQQ{{\kern 0.27em \vrule height1.45ex width0.03em depth0em
          \kern-0.30em \rm Q}}
\def\Q{{\mathchoice
        {\sQQ}
        {\sQQ}
  {\kern 0.225em \vrule height1.05ex width0.025em depth0em \kern-0.25em \rm Q}
  {\kern 0.180em \vrule height0.78ex width0.020em depth0em \kern-0.20em \rm Q}
        }}
\def\sRR{{\rm I\kern-0.16em{}R}}
\def\R{{\mathchoice
  {\sRR}
  {\sRR}
  {\rm I\kern-0.12em{}R}
  {\rm I\kern-0.10em{}R} }}
\def\sZZ{{\rm Z\kern-0.32em{}Z}}
\def\Z{{\mathchoice
  {\sZZ}
  {\sZZ} 
  {\rm Z\kern-0.3em{}Z}      
  {\rm Z\kern-0.25em{}Z} }}  
\def\ZZZ{{\rm Z\kern-0.24em{}Z}}
\def\sKK{{\rm I\kern-0.16em{}K}}
\def\K{{\mathchoice
  {\sKK}
  {\sKK}
  {\rm I\kern-0.12em{}K}
  {\rm I\kern-0.10em{}K} }}

\begin{document}
\renewcommand{\baselinestretch}{1}

\thispagestyle{empty}
{\flushright{\small MIT-CTP-2952\\hep-th/0003085\\}}

\vspace{.3in}
\begin{center}\LARGE {D-Brane Gauge Theories from Toric Singularities
	and Toric Duality}
\end{center}

\vspace{.2in}
\begin{center}
{\large Bo Feng, Amihay Hanany and Yang-Hui He\\}
\normalsize{fengb, hanany, yhe@ctp.mit.edu\footnote{
Research supported in part
by the CTP and the LNS of MIT and the U.S. Department of Energy 
under cooperative research agreement \# DE-FC02-94ER40818.
A. H. is also supported by an A. P. Sloan Foundation Fellowship,
a DOE OJI award and by the NSF under grant no. PHY94-07194.}
\\}
\vspace{.2in} {\it Center for Theoretical Physics,\\ Massachusetts
Institute of Technology\\ Cambridge, Massachusetts 02139, U.S.A.\\}
\end{center}
\vspace{0.1in}
\begin{abstract}
Via partial resolution of Abelian orbifolds we present an algorithm
for extracting a consistent set of gauge theory data for an arbitrary
toric variety whose singularity a D-brane probes.
As illustrative examples, we
tabulate the matter content and superpotential for a D-brane living on
the toric del Pezzo surfaces as well as the zeroth Hirzebruch surface.
Moreover, we discuss the non-uniqueness of the general problem and present
examples of vastly different theories whose moduli spaces are
described by the same toric data. Our methods provide new tools
for calculating gauge theories which flow to the same universality
class in the IR. We shall call it ``Toric Duality.''
\end{abstract}
\newpage

\section{Introduction}
The study of D-branes as probes of geometry and topology of
space-time has by now been of wide practice (cf. e.g. \cite{Lecture}).
In particular, the
analysis of the moduli space of gauge theories, their matter content,
superpotential and $\beta$-function, as world-volume theories of
D-branes sitting at geometrical singularities is still a widely
pursued topic. Since the pioneering work in \cite{Orb1}, where the
moduli and
matter content of D-branes probing ALE spaces had been extensively
investigated, much work ensued. The primary focus on (Abelian)
orbifold singularities of the type $\C^2/\Z_n$ was quickly
generalised using McKay's Correspondence, to arbitrary (non-Abelian)
orbifold singularities $\C^2/(\Gamma \subset SU(2))$, i.e., to
arbitrary ALE spaces, in \cite{Orb2}.

Several directions followed. With the realisation \cite{KS,Morrison} 
that these singularities provide various horizons,
\cite{Orb1,Orb2} was
quickly generalised to a treatment for arbitrary finite subgroups
$\Gamma \subset SU(N)$, i.e., to generic Gorenstein singularities, by
\cite{LNV}. The case of $SU(3)$ was then promptly studied in
\cite{Han-He,Muto,Greene2} using this technique and a generalised McKay-type
Correspondence was proposed in \cite{Han-He,He-Song}.
Meanwhile, via T-duality transformations, certain orbifold
singularities can be
mapped to type II brane-setups in the fashion of \cite{Han-Wit}. The
relevant gauge theory data on the world volume can thereby be
conveniently read from configurations of NS-branes, D-brane stacks as
well as orientifold planes. For $\C^2$ orbifolds, the $A$ and $D$
series have been thus treated \cite{Han-Wit,Kapustin}, whereas for 
$\C^3$ orbifolds, the Abelian case of $\Z_k \times \Z_{k'}$ has been 
solved by the brane box models \cite{Han-Zaf,Han-Ura}. First examples
of non-Abelian $\C^3$ orbifolds have been addressed in \cite{Bo-Han}
as well as recent works in \cite{Muto2}.

Thus rests the status of orbifold theories. What we note in particular
is that once we specify the properties of the orbifold in terms of the
algebraic properties of the finite group, the gauge theory information
is easily extracted.
Of course, orbifolds are a small subclass of algebro-geometric
singularities. This is where we move on to {\bf toric varieties}.
Inspired by the linear $\sigma$-model approach of
\cite{Witten}, which provides a rich structure of the moduli space,
especially in connexion with various geometrical phases of the
theory, the programme of utilising toric methods to study the
behaviour of the
gauge theory on D-branes which live on and hence resolve certain 
singularities was initiated in \cite{DGM}. In this light, toric
methods provide 
a powerful tool for studying the moduli space of the
gauge theory. In treating the F-flatness and D-flatness conditions for
the SUSY vacuum in conjunction, these methods show how branches of the
moduli space and hence phases of the theory may be parametrised by the
algebraic equations of the toric variety. Recent developments in
``brane diamonds,'' as an extension of the brane box rules,
have been providing great insight to such a wider class of toric
singularities, especially the generalised conifold, via blown-up
versions of the standard brane setups \cite{Karch}.
Indeed, with toric techniques much information could be
extracted as we can actually analytically describe patches of the
moduli space.

Now Abelian orbifolds have toric descriptions and the above
methodolgy is thus immediately applicable thereto. While bearing in
mind that though non-Abelian orbifolds have no toric descriptions, a single
physical D-brane has been placed on various general toric singularities.
Partial resolutions of $\C^3 / (\Z_2 \times \Z_2)$, such as the
conifold and the suspended pinched point have been investigated in
\cite{Greene,Muto3} and brane setups giving the field theory contents
are constructed by \cite{Unge,Uranga,Oh}. Groundwork for the next
family, coming from the toric orbifold $\C^3 / (\Z_3 \times \Z_3)$,
such as the del Pezzo surfaces and the zeroth Hirzebruch,
has been laid in \cite{Chris}. Essentially, given the gauge
theory data on the D-brane world volume, the procedure of transforming
this information (F and D terms) into toric data which parametrises
the classical moduli space is by now well-established.

One task is therefore immediately apparent to us: how do we proceed in
the reverse direction, i.e., {\em when we probe a toric singularity with a
D-brane, how do we know the gauge theory on its world-volume?}
We recall that in the case of orbifold theories, 
\cite{LNV} devised a general method to extract the
gauge theory data (matter content, superpotential etc.) from the
geometry data (the characters of the finite group $\Gamma$),
and {\it vice versa} given the geometry, brane-setups for example, conveniently
allow us to read out the gauge theory data. 
The same is not true for toric singularities, and the second half of
the above bi-directional convenience, namely, a general method which
allows us to treat the inverse problem of extracting gauge theory data
from toric data is yet pending, or at least not in circulation.

The reason for this shortcoming is, as we shall see later, that the
problem is highly non-unique. It is thus the purpose of this writing
to address this inverse problem: given the geometry data in terms of a
toric diagram, how does one read out (at least one) gauge theory data
in terms of the matter content and superpotential? We here present
precisely this algorithm which takes the matrices encoding the
singularity to the matrices encoding a good gauge theory of the
D-brane which probes the said singularity.

The structure of the paper is as follows. In Section 2 we review the
procedure of proceeding from the gauge theory data to the toric data,
while establishing nomenclature. In Subsection 3.1, we demonstrate how
to extract the matter content and F-terms from the charge matrix of
the toric singularity. In Subsection 3.2, we exemplify our algorithm
with the well-known suspended pinched point before presenting in
detail in Subsection 3.3, the general algorithm of how to obtain the
gauge theory information from the toric data by the method of partial
resolutions. In Subsection 3.4, we show how to integrate back to
obtain the actual superpotential once the F-flatness equations are
extracted from the toric data.
Section 4 is then devoted to the illustration of our algorithm by
tabulating the D-terms and F-terms of D-brane world volume theory on
the toric del Pezzo surfaces and Hirzebruch zero.
We finally discuss in Section 5, the non-uniqueness of the inverse
problem and provide, through the studying of two types of ambiguities, ample
examples of rather different gauge theories flowing to the same toric
data. Discussions and future prospects are dealt with in Section 6.

\section{The Forward Procedure: Extracting Toric Data From Gauge Theories}
We shall here give a brief review of the procedures involved in
going from gauge theory data on the D-brane to toric data of the
singularity, using primarily the notation and concepts from \cite{DGM}.
In the course thereof special attention will be paid on 
how toric diagrams, SUSY fields and linear
$\sigma$-models weave together.

A stack of $n$ D-brane probes on algebraic singularities gives rise to
SUSY gauge
theories with product gauge groups resulting from the projection of
the $U(n)$ theory on the original stack by the geometrical structure
of the singularity.
For orbifolds $\C^k/\Gamma$, we can use the structure of the finite
group $\Gamma$ to
fabricate product $U(n_i)$ gauge groups \cite{Orb1,Orb2,LNV}.
For toric singularities,
since we have only (Abelian) $U(1)$ toroidal actions, we are so far
restricted to product $U(1)$ gauge groups\footnote{
	Proposals toward generalisations to D-brane stacks have
	been made \cite{Chris}.}.
In physical terms, we have
{\it a single D-brane probe}. Extensive work has been done in
\cite{Chris,DGM} to see how the geometrical structure of the variety
can be thus probed and how the gauge theory moduli may be
encoded. 
The subclass of toric singularities, namely Abelian orbifolds, has been
investigated to great detail \cite{Orb1,Orb3,DGM,Muto3,Chris}
and we shall make liberal usage of their properties throughout.

\begin{figure}
\centerline{\psfig{figure=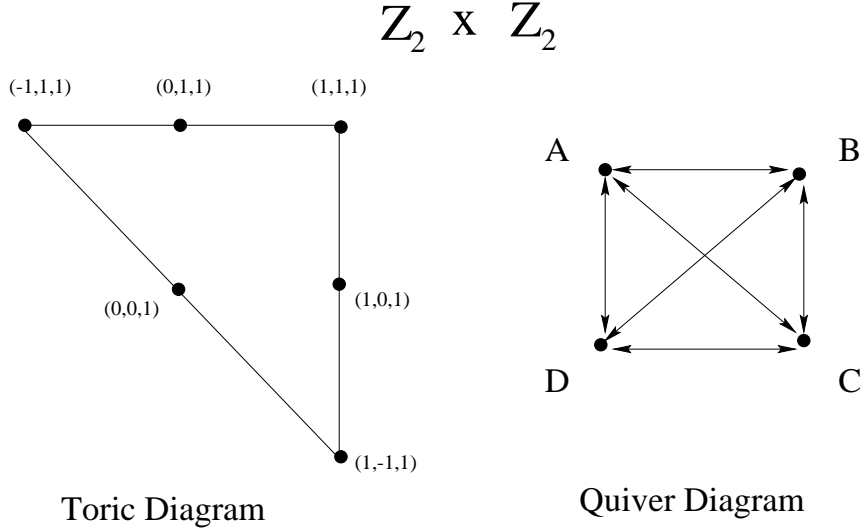,width=4.5in}}
\caption{The toric diagram for the singularity $\C^3/(\Z_2 \times
	\Z_2)$ and the quiver diagram for the gauge theory living on a
	D-brane probing it. We have labelled the nodes of the toric
	diagram by columns of $G_t$ and those of the quiver, with the
	gauge groups $U(1)_{\{A,B,C,D\}}$.}
\label{f:z2z2}
\end{figure}

Now let us consider the world-volume theory on the D-brane probe on
a toric singularity. Such a theory, as it is a SUSY gauge theory,
is characterised by its matter
content and interactions. The former is specified by quiver diagrams
which in turn give rise to {\bf D-term} equations; the latter is given
by a superpotential, whose partial derivatives with respect to the
various fields are the so-called {\bf F-term} equations. F and
D-flatness subsequently describe the (classical) moduli space of the theory.
The basic idea is that the D-term equations together with the FI-parametres,
in conjunction with the F-term equations, can be 
concatenated together into a matrix which gives the vectors forming
the dual cone of the toric variety which the D-branes probe. 
We summarise the algorithm of obtaining the toric data from the gauge
theory in the following, and to illuminate our abstraction and notation we
will use the simple example of the Abelian orbifold
$\C^3/(\Z_2 \times \Z_2)$ as given in \fref{f:z2z2}.
\begin{enumerate}
\item \underline{Quivers and D-Terms:}
	\begin{enumerate}
	\item The bi-fundamental matter content of the gauge theory
		can be conveniently
		encoded into a {\bf quiver diagram} ${\cal Q}$, which
		is simply the (possibly
		directed) graph whose {\bf adjacency matrix} $a_{ij}$
		is precisely the matrix of the
		bi-fundamentals. In the case of an Abelian
		orbifold\footnote{This is true
			for all orbifolds but of course only Abelian ones have
			known toric description.}
		prescribed by the group
		$\Gamma$, this diagram is the McKay Quiver (i.e., for
		the irreps $R_i$ 
		of $\Gamma$, $a_{ij}$ is such that $R \otimes R_i =
		\oplus_j a_{ij}R_j$
		for some fundamental representation $R$).
		We denote the set of nodes as ${\cal Q}_0 := \{v\}$
		and the set of the edges, ${\cal Q}_1 := \{a\}$.
		We let the number of nodes be $r$; for Abelian
		orbifolds, $r=|\Gamma|$ (and
		for generic orbifolds $r$ is the number of conjugacy
		classes of $\Gamma$).
		Also, we let the number of edges be $m$; this number
		depends on the number
		of supersymmetries which we have.
		The adjacency matrix (bi-fundamentals) is thus $r
		\times r$
		and the gauge group is $\prod\limits_{j=1}^r
		SU(w_j)$. For our example of $\Z_2 \times \Z_2$,
		$r=4$, 
		indexed as 4 gauge groups $U(1)_A \times U(1)_B 
		\times U(1)_C \times U(1)_D$ corresponding to the 4
		nodes, while $m=4 \times 3 = 12$,
		corresponding to the 12 arrows in \fref{f:z2z2}.
		The adjacency matrix for the quiver is
		${\tiny \left( \matrix{ 0 & 1 & 1 & 1 \cr
		1 & 0 & 1 & 1 \cr 1 & 1 & 0 & 1 \cr 1 & 1 & 1 & 0
    		\cr} \right)}$.
		Though for such simple
		examples as Abelian orbifolds and conifolds,
		brane setups and \cite{LNV} specify
		the values of $w_j$ as well as $a_{ij}$
		completely\footnote{For arbitrary
			orbifolds, $\sum\limits_j w_i n_i = |\Gamma|$
			where $n_i$ are 
			the dimensions of the irreps of $\Gamma$; for
			Abelian case, $n_i =1$.
			}, there is yet no
		discussion in the literature of obtaining the matter
		content and gauge group
		for generic toric varieties in a direct and systematic
		manner and 
		a partial purpose of this note is to present a
		solution thereof.
	\item From the $r \times r$ adjacency matrix, we construct a
		so-called $r \times m$ 
		{\bf incidence matrix} $d$ for ${\cal Q}$; this matrix is
		defined as 
		$d_{v,a} := \delta_{v,head(a)} - \delta_{v,tail(a)}$
		for $v \in {\cal Q}_0$
		and $a \in {\cal Q}_1$. Because each column of 
		$d$ must contain a 1, a $-1$ and the rest 0's by
		definition, one row
		of $d$ is always redundant; this physically signifies
		the elimination of
		an overall trivial $U(1)$ corresponding to the COM
		motion of the branes.
		Therefore we delete a row of $d$ to define the matrix
		$\Delta$ of dimensions $(r-1) \times m$; and we could
		always extract $d$ from $\Delta$ by adding a row so
		as to force each column to sum to zero. This matrix
		$\Delta$ thus contains almost as much information as
		$a_{ij}$ and once it is specified,
		the gauge group and matter content are also, with the
		exception that precise adjoints (those charged under
		the same gauge group factor and hence correspond to
		arrows that join a node to itself) are not manifest.
		For our example the $ 4 \times 12$ matrix $d$ is as
		follows and $\Delta$ is the top 3 rows:
		\[
		{\scriptsize
		d = \left(
		\begin{array}{c|cccccccccccc}
		&X_{AD}&X_{BC}&X_{CB}&X_{DA}&X_{AB}&X_{BA}&X_{CD}&
		X_{DC}&X_{AC}&X_{BD}&X_{CA}&X_{DB} \\
		A&-1&0&0&1&-1&1&0&0&-1&0&1&0 \\
		B&0&-1&1&0&1&-1&0&0&0&-1&0&1\\
		C&0&1&-1&0&0&0&-1&1&1&0&-1&0 \\ \hline 
		D&1&0&0&-1&0&0&1&-1&0&1&0&-1
		\end{array}
		\right)
		}
		\]
	\item The moment maps, arising in the sympletic-quotient
		language of the toric
		variety, are simply $\mu := d \cdot |x(a)|^2$ where
		$x(a)$ are the affine coordinates of the $\C^r$ for
		the torus $(\C^*)^r$ action.
		Physically, $x(a)$ are of course the bi-fundamentals
		in chiral multiplets (in our example they are 
		$X_{ij \in \{A,B,C,D\}}$ as labelled above)
		and the D-term equations for each
		$U(1)$ group is \cite{Witten}
		\[
		D_i=-e^2(\sum_a d_{i a} |x(a)|^2- \zeta_i)
		\]
		with $\zeta_i$ the FI-parametres. In matrix form we
		have $\Delta \cdot |x(a)|^2 = \vec{\zeta}$ and see
		that D-flatness gives precisely the moment map.
		These $\zeta$-parametres will encode the resolution of
		the toric singularity as we shall shortly see.
	\end{enumerate}
\item \underline{Monomials and F-Terms:}
	\begin{enumerate}
	\item From the super-potential $W$ of the SUSY gauge
		theory, one can write the F-Term equation
		as the system ${\partial \over {\partial X_j}} W = 0$.
		The remarkable fact is that we could solve the said
		system of equations and
		express the $m$ fields $X_i$ in terms of $r+2$ parametres
		$v_j$ which can be summarised by a matrix $K$.
		\beq
		\label{kmatrix}
		X_i =\prod_j v_j^{K_{ij}}, \qquad i=1,2,..,m;~~~j=1,2,..,r+2
		\eeq
		This matrix $K$ of dimensions $m \times (r+2)$ is the
		analogue of
		$\Delta$ in the sense that it encodes the F-terms and
		superpotential
		as $\Delta$ encodes the D-terms and the matter
		content.
		In the language of toric geometry $K$ defines a
		cone\footnote
		{
		We should be careful in this
		definition. Strictly speaking we have a lattice ${\bf
		M}=\Z^{r+2}$ with its dual lattice ${\bf N} \cong
		\Z^{r+2}$. Now let there be a set of $\Z_+$-independent
		vectors $\{\vec{k}_i\} \in {\bf M}$ and a cone is defined to
		be generated by these vectors as
		$\sigma := \{\sum_i a_i \vec{k}_i~|~a_i \in
		\R_{\ge 0}\}$; Our ${\bf M_+}$ should be ${\bf M}
		\cap \sigma$. In
		much of the literature ${\bf M_+}$ is taken to be
		simply ${\bf M'_+} := \{\sum_i a_i \vec{k}_i~|~a_i \in
		\Z_{\ge 0}\}$ in which case we must make sure that any
		lattice point contained in ${\bf M_+}$ but not in
		${\bf M'_+}$ must be counted as an independent
		generator and be added to the set of generators
		$\{\vec{k}_i\}$. After including all such points we
		would have ${\bf M'_+} = {\bf M_+}$.
		Throughout our analyses, our cone defined by $K$ as
		well the dual cone $T$ will be constituted by such a
		complete set of generators.
		}
		${\bf M_+}$ : a non-negative linear combination of $m$
		vectors $\vec{K_i}$ in an integral lattice $\Z^{r+2}$.\\
		For our example, the superpotential is
		\[
		\begin{array}{c}
		W = X_{AC}X_{CD}X_{DA} - X_{AC}X_{CB}X_{BA} + X_{CA}X_{AB}X_{BC}
		- X_{CA}X_{AD}X_{DC} \\
		+ X_{BD}X_{DC}X_{CB} - X_{BD}X_{DA}X_{AB}
		- X_{DB}X_{BC}X_{CD},
		\end{array}
		\]
		giving us 12 F-term equations and with the manifold of
		solutions parametrisable by $4+2$ new fields, whereby
		giving us the $12 \times 6$ matrix (we here show the
		transpose thereof, thus the horizontal direction
		corresponds to the original fields $X_i$ and the
		vertical, $v_j$):
		\[
		{\scriptsize
		K^t = \left(
		\begin{array}{c|cccccccccccc}
		& X_{AC} & X_{BD} & X_{CA} & X_{DB} & X_{AB} & X_{BA}
		& X_{CD} & X_{DC} & X_{AD} & X_{BC} & X_{CB} & X_{DA}
		\\ \hline
		v_1 =X_{AC} & 1 & 0 & 0 & 1 & 1 & 0 & 0 & 1 & 0 & 0 & 0 & 0 \\
		v_2 =X_{BD} & 0 & 1 & 1 & 0 & -1& 0 & 0 & -1& 0 & 0 & 0 & 0 \\
		v_3 =X_{BA} & 0 & 0 & 0 & 0 & 0 & 1 & 0 & 1 & 0 & 1 & 0 & 1 \\
		v_4 =X_{CD} & 0 & 0 & 0 & 0 & 1 & 0 & 1 & 0 & 0 & -1& 0 & -1\\
		v_5 =X_{AD} & 0 & 0 & -1& -1& 0 & 0 & 0 & 0 & 1 & 1 & 0 & 0 \\
		v_6 =X_{CB} & 0 & 0 & 1 & 1 & 0 & 0 & 0 & 0 & 0 & 0 & 1 & 1
		\end{array}
		\right).
		}
		\]
		For example, the third column reads $X_{CA} = v_2
		v_5^{-1} v_6$, i.e., $X_{AD} X_{CA} = X_{BD} X_{CB}$,
		which the the F-flatness condition ${\partial W \over
		{\partial X_{DC}}=0}$. The details of obtaining $W$
		and $K$ from each other are discussed in
		\cite{DGM,Chris} and Subsection 3.4.
	\item We let $T$ be the space of (integral) vectors dual to
		$K$, i.e., $K \cdot T \ge 0$ for all entries; this gives an
		$(r+2) \times c$ matrix for some positive integer
		$c$. Geometrically, this is the definition of a dual
		cone ${\bf N_+}$ composed of vectors $\vec{T_i}$
		such that $\vec{K} \cdot \vec{T} \ge 0$. The physical
		meaning for doing so is that $K$ may have negative
		entries which may give rise to unwanted singularities
		and hence we define a new set of $c$ fields
		$p_i$ ({\it a priori} we do not know the number $c$
		and we present the standard algorithm of finding dual
		cones in the Appendix). Thus we reduce (\ref{kmatrix})
		further into
		\beq
		\label{p_i}
			v_j=\prod_{\alpha}p_{\alpha}^{T_{j\alpha}}
		\eeq
		whereby giving $X_i = 
		\prod_j v_j^{K_{ij}}= \prod_{\alpha} 
		p_{\alpha}^{\sum_j K_{ij} T_{j\alpha}}$ with 
		$\sum_j K_{ij} T_{j\alpha}\geq 0$.
		For our $\Z_2 \times Z_2$ example, $c=9$ and
		\[
		{\scriptsize
		T_{j\alpha} = \left(
		\begin{array}{c|ccccccccc}
		& p_1 & p_2 & p_3 & p_4 & p_5 & p_6 & p_7
			& p_8 & p_9 \\ \hline
		X_{AC} & 1 & 1 & 0 & 0 & 0 & 0 & 0 & 0 & 1 \cr
		X_{BD} & 0 & 1 & 1 & 0 & 0 & 0 & 0 & 0 & 1 \cr
		X_{BA} & 0 & 0 & 1 & 1 & 1 & 0 & 0 & 0 & 0 \cr
		X_{CD} & 0 & 0 & 1 & 0 & 1 & 1 & 0 & 0 & 0 \cr
		X_{AD} & 0 & 0 & 0 & 0 & 0 & 1 & 1 & 0 & 1 \cr
		X_{CB} & 0 & 0 & 0 & 0 & 0 & 1 & 1 & 1 & 0
		\end{array}
		\right)
		}
		\]
	\item These new variables $p_\alpha$ are the
		matter fields in Witten's linear $\sigma$-model.
		How are these fields charged? We have
		written $r+2$ fields $v_j$ in terms of $c$ fields
		$p_\alpha$, and hence need $c-(r+2)$ relations to reduce
		the independent variables. Such a reduction can be
		done via the introduction of the new gauge group
		$U(1)^{c-(r+2)}$ acting on the $p_i$'s so as to give a
		new set of D-terms. The charges of these fields can be
		written as $Q_{k \alpha}$. The gauge invariance condition of
		$v_i$ under $U(1)^{c-(r+2)}$, by (\ref{p_i}), demands
		that the $(c - r - 2) \times c$ matrix
		$Q$ is such that $\sum_{\alpha}
		T_{j\alpha}Q_{k\alpha}=0$. This then defines for us our
		charge matrix $Q$ which is the cokernel of $T$:
		\[
		TQ^t=(T_{j \alpha})(Q_{k\alpha})^t = 0, \qquad
			j=1,..,r+2;~~\alpha=1,..,c;~~k=1,..,(c-r-2)
		\]
		For our example, the charge matrix is $(9-4-2) \times
		9$ and one choice is
		${\scriptsize Q_{k \alpha} = \left( 
		\matrix{
		0 & 0 & 0 & 1 & -1& 1 & -1& 0 & 0 \cr
		0 & 1 & 0 & 0 & 0 & 0 & 1 & -1& -1 \cr
		1 &-1 & 1 & 0 & -1& 0 & 0 & 0 & 0 \cr }
		\right)}$.
	\item In the linear $\sigma$-model language, the F-terms and
		D-terms can be treated in the same footing, i.e., as
		the D-terms (moment map) of the new fields $p_\alpha$; with the
		crucial difference being that the former must be set
		exactly to zero\footnote{Strictly speaking, we could
			have an F-term set to a non-zero constant. An example
			of this situation could be when there is a term $a
			\phi + \phi \tilde{Q} Q$ in the superpotential for
			some chargeless field $\phi$ and charged fields
			$\tilde{Q}$ and $Q$. The F-term for $\phi$
			reads $\tilde{Q} Q = -
a$ and not 0. However, in
			our context $\phi$ behaves like an integration
			constant and
			for our purposes, F-terms are set exactly to zero.}
		while the latter are to be resolved by
		arbitrary FI-parameters.\\
		Therefore in addition to finding the charge matrix $Q$
		for the new fields $p_\alpha$ coming from the original
		F-terms as done above, we must also find the
		corresponding charge matrix $Q_D$ for the $p_i$ coming from
		the original D-terms.
		We can find $Q_D$ in two steps. Firstly, we know the
		charge matrix for $X_i$ under $U(1)^{r-1}$, which is
		$\Delta$. By (\ref{kmatrix}), we transform the charges
		to that of the $v_j$'s, by introducing an $(r-1) \times
		(r+2)$ matrix $V$ so that $V \cdot K^t =
		\Delta$. To see this, let the charges of $v_j$ be
		$V_{lj}$ then by (\ref{kmatrix}) we have $\Delta_{li}
		= \sum\limits_j V_{lj} K_{ij} = V \cdot K^t$.
		A convenient $V$ which does so for our $\Z_2 \times
                \Z_2$ example is
                ${\scriptsize \left( 
                \matrix{
                1 & 0 & -1& 0 & 1 & 0 \cr
                0 & 1 & 1 & 0 & 0 & -1\cr
                -1& 0 & 0 & 1 & 0 & 1 \cr}
                \right)_{(4-1)\times (4+2)}}$.
		Secondly, we use (\ref{p_i}) to transform the
		charges from $v_j$'s to our final variables $p_\alpha$'s,
		which is done by introducing an
		$(r + 2) \times c$ matrix $U_{j \alpha}$ so that
                $U \cdot T^t = {\rm Id}_{(r+2) \times (r+2)}$. In our
		example, one choice for $U$ is
                ${\scriptsize U_{j \alpha} = \left( 
                \matrix{
                1 & 0 & 0 & 0 & 0 & 0 & 0 & 0 & 0 \cr
                -1& 1 & 0 & 0 & 0 & 0 & 0 & 0 & 0 \cr
                0 & 0 & 0 & 1 & 0 & 0 & 0 & 0 & 0 \cr
                0 & 0 & 0 & 0 & 0 & 1 & -1& 0 & 0 \cr
                0 & -1& 0 & 0 & 0 & 0 & 0 & 0 & 1 \cr
                0 & 0 & 0 & 0 & 0 & 0 & 0 & 1 & 0 }
                \right)_{(4+2) \times 9}}$.\\
		Threfore, combining the two steps, we obtain $Q_D = V \cdot
		U$ and in our example, ${\scriptsize (V \cdot U)_{l
		\alpha} = \left(
		\matrix{ 1 & -1& 0 & -1& 0 & 0 & 0 & 0 & 1 \cr
			-1& 1 & 0 & 1 & 0 & 0 & 0 & -1& 0 \cr
			-1& 0 & 0 & 0 & 0 & 1 & -1& 1 & 0} \right)}$.
	\end{enumerate}
\item Thus equipped with the information from the two sides: the
	F-terms and D-terms, and with the two required charge matrices
	$Q$ and  $V \cdot U$ obtained, finally we concatenate them to
	give a $(c-3) \times c$ matrix $Q_t$.
	The transpose of the kernel of $Q_t$, with (possible
	repeated columns) gives rise to a matrix $G_t$.
	The columns of this resulting $G_t$
	then define the vertices of the toric diagram
	describing the polynomial corresponding to the singularity on which
	we initially placed our D-branes.
	Once again for our example,
	$Q_t = {\scriptsize \left(
	\begin{array}{ccccccccc|c}
	0 & 0 & 0 & 1 & -1& 1 & -1& 0 & 0 &  0\cr
	0 & 1 & 0 & 0 & 0 & 0 & 1 & -1& -1&  0 \cr
	1 & -1& 1 & 0 & -1& 0 & 0 & 0 & 0 &  0 \cr \hline
	1 & -1& 0 & -1& 0 & 0 & 0 & 0 & 1 & \zeta_1\cr
	-1& 1 & 0 & 1 & 0 & 0 & 0 & -1& 0 & \zeta_2\cr
	-1& 0 & 0 & 0 & 0 & 1 & -1& 1 & 0 & \zeta_3
	\end{array}
	\right)}$ and
	$G_t = {\scriptsize \left(
	\matrix{
	0 & 1 & 0 & 0 & -1& 0 & 1 & 1 & 1 \cr
	1 & 1 & 1 & 0 & 1 & 0 & -1& 0 & 0 \cr
	1 & 1 & 1 & 1 & 1 & 1 & 1 & 1 & 1}
	\right)}$. The columns of $G_t$, up to repetition, are
	precisely marked in the toric diagram for $\Z_2 \times \Z_2$
	in \fref{f:z2z2}.
\end{enumerate}

Thus we have gone from the F-terms and the D-terms of the gauge theory
to the nodes of the toric diagram. In accordance with
\cite{Fulton}, $G_t$ gives the algebraic variety whose equation is
given by the maximal ideal in the polynomial ring \\
$\C[YZ,XYZ,Z,X^{-1}YZ,XY^{-1}Z,XZ]$ (the exponents $(i,j,k)$
in $X^iY^jZ^k$ are exactly the columns), which is $uvw=s^2$, upon
defining $u=(YZ)(XYZ)^2(Z)(XZ)^2; v=(YZ)^2(Z)^2(X^{-1}YZ)^2;
w=(Z)^2(XY^{-1}Z)(XZ)^2$ and \\$s = (YZ)^2 (XYZ) (Z)^2 (X^{-1}YZ)
(XY^{-1}Z) (XZ)^2$; this is precisely $\C^3/(\Z_2 \times \Z_2)$.
In physical terms this equation
parametrises the moduli space obtained from the F and D flatness of
the gauge theory.

We remark two issues here. In the case of there being no
superpotential we could still define $K$-matrix. In this case, with
there being no F-terms, we simply take $K$ to be the identity. This
gives $T=$Id and $Q=0$. Furthermore $U$ becomes Id and $V=\Delta$,
whereby making $Q_t = \Delta$ as expected because all information
should now be contained in the D-terms.
Moreover, we note that the very reason we can construct a $K$-matrix
is that all of the equations in the F-terms we deal with are in the
form $\prod\limits_i X_i^{a_i} = \prod\limits_j X_j^{b_j}$; this
holds in general if every field $X_i$ appears twice and precisely twice in the
superpotential. More generic situations would so far transcend the
limitations of toric techniques.

Schematically, our procedure presented above at length, what it
means is as follows: we begin with
two pieces of physical data: (1) matrix $d$ from the quiver encoding
the gauge groups and D-terms and (2) matrix $K$ encoding the F-term
equations. From these we extract the matrix $G_t$ containing the toric
data by the flow-chart:
\[
\begin{array}{ccccccc}
\mbox{Quiver} \rightarrow d	& \rightarrow	&\Delta	& & & & \\
	&	&\downarrow	&	&	&	&	\\
\mbox{F-Terms} \rightarrow K	& \stackrel{V \cdot K^t =
	\Delta}{\rightarrow}
		& V	 & & & & \\
\downarrow	&	& \downarrow	& & & & \\
T = {\rm Dual}(K)	& \stackrel{U \cdot T^t = {\rm
	Id}}{\rightarrow} & U & \rightarrow & VU & &\\
\downarrow	&	&	&	& \downarrow	& & \\
Q = [{\rm Ker}(T)]^t	&	& \longrightarrow	& & Q_t =
	\left( \begin{array}{c}
					Q \\ VU \end{array} \right) &
						\rightarrow & G_t =
	[{\rm Ker}(Q_t)]^t \\

\end{array}
\]


\section{The Inverse Procedure: Extracting Gauge Theory Information
	from Toric Data}
As outlined above we see that wherever possible, the gauge theory of a D-brane
probe on certain singularities such as Abelian orbifolds, conifolds, etc., can
be conveniently encoded into the matrix $Q_t$ which essentially concatenates
the information contained in the D-terms and F-terms of the original gauge theory.
The cokernel of this matrix is then a list of vectors which prescribes the toric
diagram corresponding to the singularity.
It is natural to question ourselves whether the converse could be done, i.e.,
whether given an arbitrary singularity which affords a toric description, we could
obtain the gauge theory living on the D-brane which probes the said singularity.
This is the inverse problem we projected to solve in the introduction.

\subsection{Quiver Diagrams and F-terms from Toric Diagrams}
Our result must be two-fold: first, we must be able to extract the
D-terms, or
in other words the quiver diagram which then gives the gauge group and
matter content; second, we must extract the F-terms, which we can
subsequently integrate back to give the superpotential. These two
pieces of data then suffice to specify the gauge theory.
Essentially we wish to trace the arrows in the above flow-chart from
$G_t$ back to $\Delta$ and $K$. The general methodology seems
straightforward:
\begin{enumerate}
\item Read the column-vectors describing the nodes of the given toric
	diagram, repeat the
	appropriate columns to obtain $G_t$ and then set $Q_t={\rm
	Coker}(G_t)$; 
\item Separate the D-term ($V \cdot U$) and F-term ($Q_t$) portions
	from $Q_t$;
\item From the definition of $Q$, we obtain\footnote{As
		mentioned before we must ensure that such a $T$ be chosen
		with a complete set of $\Z_+$-independent generators;}
	$T = \ker (Q)$.
\item Farka's Theorem \cite{Fulton} guarantees that the dual of a
	convex polytope remains convex whence we could invert and have
	$K = {\rm Dual}(T^t)$; Moreover the duality theorem gives that
	Dual(Dual($K)) = K$, thereby facilitating the inverse procedure. 
\item Definitions $U \cdot T^t = \rm{Id}$ and 
	$V \cdot K^t = \Delta$ $\Rightarrow$
	$(V \cdot U) \cdot (T^t \cdot K^t) = \Delta$.
\end{enumerate}

We see therefore that once the appropriate $Q_t$ has been found, the relations
\beq
\label{Delta}
K= {\rm Dual}(T^t) \qquad \Delta = (V \cdot U) \cdot (T^t \cdot K^t)
\eeq
retrieve our desired $K$ and $\Delta$.
The only setback of course is that the appropriate $Q_t$
is NOT usually found.
Two ambiguities are immediately apparent to us:
(A) In step 1 above, there is really no way to know a priori which of
	the vectors we should repeat when writing into the $G_t$ matrix;
(B) In step 2, to separate the D-terms and the F-terms, i.e.,
	which rows constitute $Q$ and which constitute $V \cdot U$
	within $Q_t$, seems arbitrary.
We shall in the last section discuss these ambiguities in more detail and
actually perceive it to be a matter of interest. Meanwhile, in light thereof,
we must find an alternative, to find a canonical method which avoids
such ambiguities and gives us a consistent gauge theory which has such
well-behaved properties as having only bi-fundamentals etc.; this is where
we appeal to partial resolutions.

Another reason for this canonical method is compelling. The astute
reader may question as to how could we guarantee, in our mathematical
excursion of performing the inverse procedure, that the gauge theory
we obtain at the end of the day is one that still lives on the
world-volume of a D-brane probe? Indeed, if we na\"{\i}vely traced back the
arrows in the flow-chart, bearing in mind the said ambiguities, we
have no {\it a fortiori} guarantee that we have a brane theory at
all. However, the method via partial resolution of Abelian orbifolds
(which are themselves toric) does give us assurance. When we are
careful in tuning the FI-parametres so as to stay inside
cone-partitions of the space of these parametres (and avoid flop
transitions) we do still have the resulting theory being physical
\cite{Chris}. Essentially this means that
with prudence we tune the FI-parametres in the allowed domains from
a parent orbifold theory, thereby giving a subsector theory which
still lives on the D-brane probe and is well-behaved.
Such tuning we shall practice in the following.

The virtues of this appeal to resolutions
are thus twofold: not only do we avoid ambiguities, we are further
endowed with physical theories. Let us thereby present this canonical
mathod.

%
\subsection{A Canonical Method: Partial Resolutions of Abelian Orbifolds}
%
Our programme is standard \cite{Chris}: theories on the Abelian orbifold
singularity of the form $\C^k / \Gamma$ for 
$\Gamma(k,n) = \Z_n \times \Z_n \times ... \Z_n$ ($k-1$ times) are well
studied. The complete information (and in
particular the full $Q_t$ matrix) for $\Gamma(k,n)$ is well known: $k=2$ is
the elliptic model, $k=3$, the Brane Box, etc.
In the toric context, $k=2$ has been analysed in great detail by
\cite{Orb1}, $k=3,n=2$ in e.g. \cite{Unge,Uranga,Oh}, $k=3,n=3$ in
\cite{Chris}. Now we know that given any toric diagram of
dimension $k$, we can embed it into such a $\Gamma(k,n)$-orbifold for
some sufficiently large $n$; and we choose the smallest such $n$ which
suffices. This embedding is always possible because the toric diagram
for the latter is the $k$-simplex of length $n$ enclosing lattice
points and any toric diagram, being a collection of lattice points,
can be obtained therefrom via deletions of a subset of points.
This procedure is known torically as {\bf partial resolutions} of $\Gamma(k,n)$.
The crux of our algorithm is that the deletions in the toric diagram
corresponds to the turning-on of the FI-parametres, and which in turn
induces a method to determine a $Q_t$ matrix for our original
singularity from that of $\Gamma(n,k)$.

\begin{figure}
\centerline{\psfig{figure=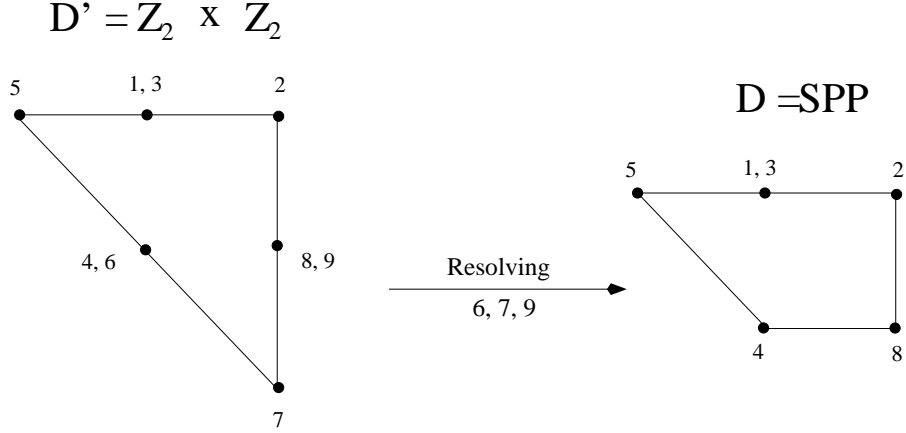,width=4.7in}}
\caption{The toric diagram showing the resolution of the 
	$\C^3 / (\Z_2 \times \Z_2)$ singularity to the suspended pinch
	point (SPP). The numbers $i$ at the nodes refer to the $i$-th
	column of the matrix $G_t$ and physically correspond 
	to the fields $p_i$ in the linear $\sigma$-model.}
\label{f:SPP}
\end{figure}

We shall first turn to an illustrative example of the suspended
pinched point singularity (SPP) and then move on to discuss generalities.
The SPP and conifold as resolutions of $\Gamma(3,2)=\Z_2 \times \Z_2$
have been extensively studied
in \cite{Uranga}. The SPP, given by $xy = zw^2$, can be obtained from
the $\Gamma(3,2)$
orbifold, $xyz = w^2$, by a single $\P^1$ blow-up. This is shown torically in
\fref{f:SPP}. Without further ado let us demonstrate our procedure.
\begin{enumerate}
\item \underline{Embedding into $\Z_2 \times \Z_2$:}
	Given the toric diagram $D$ of SPP, we recognise that it can be
	embedded minimally into the diagram $D'$ of $\Z_2 \times
	\Z_2$. Now information on $D'$ is readily at hand
	\cite{Uranga}, as presented
	in the previous section. Let us re-capitulate:
	\[
	{\scriptsize
	Q'_t := \left(
	\matrix{p_1 & p_2 & p_3 & p_4 & p_5 & p_6 & p_7 & p_8 & p_9 & \cr
		0 & 0 & 0 & 1 & -1 & 1 & -1 & 0 & 0 & 0 \cr
		0 & 1 & 0 & 0 & 0 & 0 & 1 & -1 & -1 & 0 \cr
		1 & -1 & 1 & 0 & -1 & 0 & 0 & 0 & 0 & 0 \cr
		1 & -1 & 0 & -1 & 0 & 0 & 0 & 0 & 1 & \zeta_1 \cr
		-1 & 1 & 0 & 1 & 0 & 0 & 0 & -1 & 0 & \zeta_2 \cr
		-1 & 0 & 0 & 0 & 0 & 1 & -1 & 1 & 0 & \zeta_3 \cr}
	\right),
	}
	\]
	and
	\[
	{\scriptsize
	G'_t := {\rm coker}(Q'_t) = \left(
	\matrix{p_1 & p_2 & p_3 & p_4 & p_5 & p_6 & p_7 & p_8 & p_9 \cr
		0 & 1 & 0 & 0 & -1 & 0 & 1 & 1 & 1 \cr
		1 & 1 & 1 & 0 & 1 & 0 & -1 & 0 & 0 \cr
		1 & 1 & 1 & 1 & 1 & 1 & 1 & 1 & 1 \cr}
	\right),
	}
	\]
	which is drawn in \fref{f:z2z2}. The fact that the last row of
	$G_t$ has the same number (i.e., these three-vectors are all
	co-planar) ensures that $D'$ is Calabi-Yau \cite{Lecture}.
	Incidentally, it would be very helpful for one to catalogue the
	list of $Q_t$ matrices of $\Gamma(3,n)$ for $n = 2,3...$ which
	would suffice for all local toric singularities of Calabi-Yau
	threefolds.\\
	In the above definition of $Q'_t$ we have included an extra column
	$(0,0,0,\zeta_1,\zeta_2,\zeta_3)$ so as to specify that the
	first three rows of $Q'_t$ are F-terms (and hence exactly
	zero) while the last three rows are D-terms
	(and hence resolved by FI-parametres $\zeta_{1,2,3}$). We
	adhere to the notation in \cite{Uranga} and label the columns
	(linear $\sigma$-model fields) as $p_1 ... p_9$;
	this is shown in \fref{f:SPP}.
\item \underline{Determining the Fields to Resolve by Tuning $\zeta$:}
	We note that if we turn on a single FI-parametre we
	would arrive at the SPP; this is the resolution of $D'$ to
	$D$. The subtlety is that one may need to eliminate more than
	merely the 7th column as there is more than one field
	attributed to each node in the toric diagram and eliminating
	column 7 some other columns corresponding to the adjacent
	nodes (namely out of 4,6,8 and 9) may also be eliminated.
	We need a judicious choice of $\zeta$ for a consistent blowup.
	To do so we must solve for fields $p_{1,..,9}$ and tune the
	$\zeta$-parametres such that 
	at least $p_7$ acquires non-zero VEV (and whereby resolved).
	Recalling that the D-term equations are actually linear
	equations in the modulus-squared of the fields, we shall
	henceforth define $x_i := |p_i|^2$ and consider linear-systems
	therein. 
	Therefore we perform Gaussian row-reduction on $Q'$ and solve
	all fields in terms of $x_7$ to give:
	$\vec{x} = \{x_1,x_2,x_1 + \zeta_2 + \zeta_3,
	{{2\,x_1 - x_2 + x_7 - \zeta_1 + \zeta_2}\over 2},
	2\,x_1 - x_2 + \zeta_2 + \zeta_3,
	{{2\,x_1 - x_2 + x_7 + \zeta_1 + \zeta_2 + 2\,\zeta_3}\over 2},x_7,
	{{x_2 + x_7 - \zeta_1 - \zeta_2}\over 2},
	{{x_2 + x_7 + \zeta_1 + \zeta_2}\over 2} \}$.\\
	The nodes far away from $p_7$ are clearly unaffected by the
	resolution, thus the fields corresponding thereto continue to have
	zero VEV. This means we solve the above set of solutions
	$\vec{x}$ once again, setting $x_{5,1,3,2} = 0$, with
	$\zeta_{1,2,3}$ being the variables, giving upon back
	substitution,
	$\vec{x} = \{0, 0, 0, {{x_7 - \zeta_1 - \zeta_3}\over 2}, 0, {{x_7 + 
	\zeta_1 + \zeta_3}\over 2}, x_7, {{x_7 - \zeta_1 + \zeta_3}\over 2},\\
	{{x_7 + \zeta_1 - \zeta_3}\over 2} \}$. Now we have an arbitrary
	choice and we set $\zeta_3 = 0$ and $x_7 = \zeta_1$ to make
	$p_4$ and $p_8$ have zero VEV. This makes $p_{6,7,9}$ our candidate
	for fields to be resolved and seems perfectly reasonable
	observing \fref{f:SPP}. The constraint on our choice is that
	all solutions must be $\ge 0$ (since the $x_i$'s are VEV-squared).
\item \underline{Solving for $G_t$:}
	We are now clear what the resolution requires of us: in order to
	remove node $p_7$ from $D'$ to give the SPP, 
	we must also resolve 6, 7 and 9.
	Therefore we immediately obtain $G_t$ by
	directly removing the said columns from $G'_t$:
	\[
	{\scriptsize
	G_t := {\rm coker}(Q_t) = \left(
	\matrix{p_1 & p_2 & p_3 & p_4 & p_5 & p_8 \cr 
		0 & 1 & 0 & 0 & -1 & 1 \cr
		1 & 1 & 1 & 0 & 1 & 0 \cr
		1 & 1 & 1 & 1 & 1 & 1  \cr}
	\right),
	}
	\]
	the columns of which give the toric diagram $D$ of the SPP, as
	shown in \fref{f:SPP}.
\item \underline{Solving for $Q_t$:}
	Now we must perform linear combination on the rows of $Q'_t$ to
	obtain $Q_t$ so as to force columns 6, 7 and 9 zero.
	The following constraints must be born in mind. 
	Because $G_t$ has 6 columns and
	3 rows and is in the null space of $Q_t$, which itself must
	have $9-3$ columns (having eliminated $p_{6,7,9}$), we must
	have $6-3=3$ rows for $Q_t$. Also, the row
	containing $\zeta_1$ must be eliminated as this is precisely 
	our resolution chosen above (we recall that the FI-parametres
	are such that $\zeta_{2,3} = 0$ and are hence unresolved,
	while $\zeta_1 > 0$ and must be removed from the D-terms for
	SPP).\\
	We systematically proceed. Let there be variables
	$\{a_{i=1,..,6}\}$ so that $y := \sum_i a_i {\rm
	row}_i(Q'_t)$
	is a row of $Q_t$. Then (a) the 6th, 7th and 9th
	columns of $y$ must be set to 0 and moreover (b) with these
	columns removed $y$ must be in the nullspace spanned by the
	rows of $G_t$. We note of course that since $Q'_t$ was in the
	nullspace of $G'_t$ initially, that the operation of
	row-combinations is closed
	within a nullspace, and that the columns to be set to 0 in
	$Q'_t$ to give $Q_t$ are precisely those removed in $G'_t$ to
	give $G_t$, condition (a) automatically implies (b).
	This condition (a) translates to the equations
	$\{a_1+a_6=0,-a_1+a_2-a_6=0,-a_2+a_4=0 \}$ which
	afford the solution $a_1 = -a_6; a_2=a_4=0$. The fact that
	$a_4=0$ is comforting, because it eliminates the row containing
	$\zeta_1$. We choose $a_1 = 1$. Furthermore
	we must keep row 5 as $\zeta_2$ is yet unresolved
	(thereby setting $a_5 = 1$).
	This already gives two of the 3 anticipated rows of $Q_t$: row$_5$ and
	row$_1$ - row$_6$. The remaining row must corresponds to an
	F-term since we have exhausted the D-terms, this we choose
	to be the only remaining variable: $a_3 = 1$.
	Consequently, we arrive at the matrix
	\[
	{\scriptsize
	Q_t = \left(
	\matrix{p_1 & p_2 & p_3 & p_4 & p_5 & p_8 & \cr
		1 & -1 & 1 & 0 & -1 & 0 & 0 \cr
		-1 & 1 & 0 & 1 & 0 &  -1 & \zeta_2 \cr
		-1 & 0 & 0 & -1 & 1 & 1 & \zeta_3 \cr }		
	\right).
	}
	\]
\item \underline{Obtaining $K$ and $\Delta$ Matrices:}
	The hard work is now done. We now recognise from $Q_t$ that
	$Q = (1,-1,1,0,-1,0)$, giving
	\[{\scriptsize
	T_{j \alpha} := {\rm ker}(Q) = \left(
	\matrix{0 & 0 & 0 & 0 & 0 & 1 \cr
		1 & 0 & 0 & 0 & 1 & 0 \cr
		0 & 0 & 0 & 1 & 0 & 0 \cr
		-1 & 0 & 1 & 0 & 0 & 0 \cr
		1 & 1 & 0 & 0 & 0 & 0 \cr}
	\right);
	\qquad
	K^t := {\rm Dual}(T^t) = \left(
	\matrix{1 & 0 & 0 & 0 & 0 & 0 \cr
		0 & 0 & 1 & 0 & 1 & 0 \cr
		0 & 1 & 0 & 0 & 0 & 0 \cr
		0 & 0 & 1 & 1 & 0 & 0 \cr
		0 & 0 & 0 & 1 & 0 & 1 \cr}
	\right).
	}\]	
	Subsequently we obtain
	${\tiny
	T^t \cdot K^t = \left(
	\matrix{0 & 0 & 0 & 0 & 1 & 1 \cr
		0 & 0 & 0 & 1 & 0 & 1 \cr
		0 & 0 & 1 & 1 & 0 & 0 \cr
		0 & 1 & 0 & 0 & 0 & 0 \cr
		0 & 0 & 1 & 0 & 1 & 0 \cr
		1 & 0 & 0 & 0 & 0 & 0 \cr}
	\right),
	}$
	which we do observe indeed to have every entry positive semi-definite.
	Furthermore we recognise from $Q_t$ that
	${\tiny
	V \cdot U = \left(
	\matrix{-1 & 1 & 0 & 1 & 0 & -1 \cr
		-1 & 0 & 0 & -1 & 1 & 1 \cr}
	\right),
	}$
	whence we obtain at last, using (\ref{Delta}),
	\[
	{\scriptsize
	\Delta = \left(
	\matrix{-1 & 1 & 0 & 1 & -1 & 0 \cr
		1 & -1 & 1 & 0 & 0 & -1 \cr}
	\right)
	\qquad \Rightarrow \qquad
	d = \left(
	\begin{array}{c|cccccc}
			& X_1 & X_2 & X_3 & X_4 &  X_5 & X_6 \\
		U(1)_A & -1 & 1 & 0 & 1 & -1 & 0 \cr
		U(1)_B & 1 & -1 & 1 & 0 & 0 & -1 \cr \hline
		U(1)_C & 0 & 0 & -1 & -1 & 1 & 1 \end{array}
	\right),
	}
	\]
	giving us the quiver diagram (included in \fref{f:SPPquiver}
	for reference), matter content and gauge group
	of a D-brane probe on SPP in agreement with \cite{Uranga}. We
	shall show in the ensuing sections that the superpotential we
	extract has similar accordance.
\end{enumerate}
\begin{figure}
\centerline{\psfig{figure=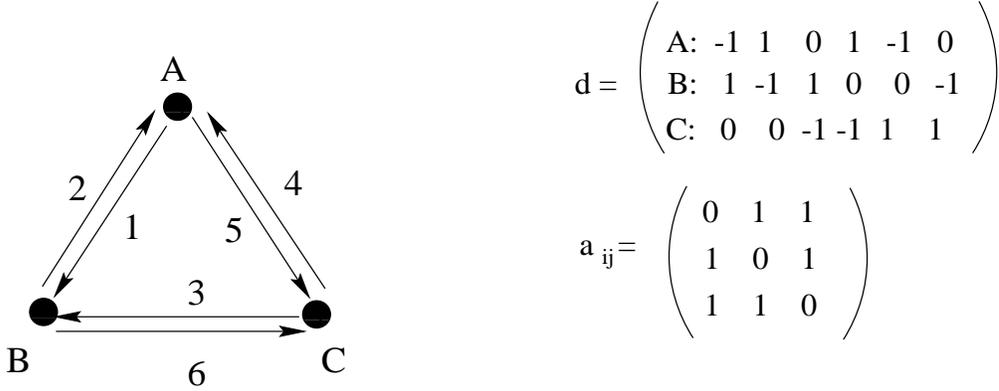,width=5.5in}}
\caption{The quiver diagram showing the matter content of a D-brane
	probing the SPP singularity. We have not marked in the
	chargeless field $\phi$ (what in a non-Abelian theory would
	become an adjoint) because thus far the toric techniques do not
	yet know how to handle such adjoints.}
\label{f:SPPquiver}
\end{figure}

\subsection{The General Algorithm for the Inverse Problem}
Having indulged ourselves in this illustrative example of the SPP, we
proceed to outline the general methodology of obtaining the gauge
theory data from the toric diagram.
\begin{enumerate}
\item \underline{Embedding into $\C^k / (\Z_n)^{k-1}$:}
	We are given a toric diagram $D$ describing an algebraic variety of
	complex dimension $k$ (usually we are concerned with local
	Calabi-Yau singularities of $k=2,3$ so that branes living
	thereon give ${\cal N} = 2,1$ gauge theories). We immediately 
	observe that $D$ could always be embedded into $D'$,
	the toric diagram of the orbifold 
	$\C^k / (\Z_n)^{k-1}$ for some sufficiently large integer $n$.
	The matrices $Q'_t$ and $G'_t$ for $D'$ are standard.
	Moreover we know that the matrix $G_t$ for our original variety
	$D$ must be a submatrix of $G'_t$. Equipped with $Q'_t$ and  $G'_t$
	our task is to obtain $Q_t$; and as an additional check we
	could verify that $Q_t$ is indeed in the nullspace of $G_t$.
\item \underline{Determining the Fields to Resolve by Tuning $\zeta$:}
	$Q'_t$ is a $k \times a$ matrix\footnote{We henceforth
		understand that there is
		an extra column of zeroes and $\zeta$'s.}
	(because $D'$ and $D$ are dimension $k$) for some
	$a$; $G'_t$, being its nullspace, is thus $(a-k) \times a$.
	$D$ is a partial resolution of $D'$. 
	In the SPP example above, we performed a single resolution
	by turning on one FI-parametre, generically however, we could
	turn on as many $\zeta$'s as the embedding permits.
	Therefore we let $G_t$ be $(a-k) \times (a-b)$
	for some $b$ which depends on the number of resolutions.
	Subsequently the $Q_t$ we need is $(k-b) \times (a-b)$.\\
	Now $b$ is determined directly by examining $D'$ and $D$;
	it is precisely the number of fields $p$ associated to those
	nodes in $D'$ we wish to eliminate to arrive at $D$.
	Exactly which $b$ columns are to be eliminated is determined
	thus: we perform Gaussian row-reduction on $Q'_t$ so as to
	solve the $k$ linear-equations in $a$ variables $x_i :=
	|p_i|^2$, with F-terms set to 0 and D-terms to FI-parametres.
	The $a$ variables are then expressed in terms of the
	$\zeta_i$'s and the set $B$ of $x_i$'s corresponding to the nodes
	which we definitely know will disappear as we resolve $D'
	\rightarrow D$. The subtlety is that in eliminating $B$, some
	other fields may also acquire non zero VEV and be eliminated;
	mathematically this means that Order$(B) < b$.\\
	Now we make a judicious choice of which fields will remain and
	set them to zero and impose this further on the solution
	$x_{i=1,..,a} = x_i(\zeta_j;B)$ from above until Order$(B) =
	b$, i.e., until we have found all the fields we need to
	eliminate. We know this occurs and that our choice was correct
	when all $x_i \ge 0$ with those equaling 0 corresponding to
	fields we do not wish to eliminate as can be observed from the
	toric diagram. If not, we modify our initial choice and repeat
	until satisfaction. This procedure then determines the $b$
	columns which we wish to eliminate from $Q'_t$.
\item \underline{Solving for $G_t$ and $Q_t$:}
	Knowing the fields to eliminate, we must thus perform linear
	combinations on the $k$ rows of $Q'_t$
	to obtain the $k-b$ rows of $Q_t$ based upon the two constraints
	that (1) the $b$ columns must be all reduced to zero 
	(and thus the nodes can be removed) and that
	(2) the $k-b$ rows (with $b$ columns removed) are in the
	nullspace of $G_t$. As mentioned in our SPP example, condition
	(1) guarantees (2) automatically.\\
	In other words, we need to solve for $k$ variables 
	$\{ x_{i = 1,..,k} \}$ such that
	\beq
	\label{elimcol}
	\sum\limits_{i=1}^k x_i~(Q'_t)_{ij} = 0 
		\qquad {\rm for}~~ j=p_1, p_2,  ... p_b \in B.
	\eeq
	Moreover, we immediately obtain $G_t$ by eliminating the $b$
	columns from $G'_t$. Indeed, as discussed earlier,
	(\ref{elimcol}) implies that
	$\sum\limits_{i=1}^k \sum\limits_{j \ne p_{1...b}} 
	x_i~(Q'_t)_{ij}~(G_t)_{mj} = 0$ for $m=1,...,a-k$ and hence
	guarantees that the $Q_t$ we obtain is in the nullspace of $G_t$.\\
	We could phrase equation (\ref{elimcol}) for $x_i$ in matrix
	notation and directly evaluate
	\beq
	\label{qt}
	Q_t = NullSpace(W)^t \cdot \tilde{Q'_t}
	\eeq
	where $\tilde{Q'_t}$ is $Q'_t$ with the appropriate columns
	($p_{1...b}$) removed and $W$ is the matrix constructed from
	the deleted columns.
\item \underline{Obtaining the $K$ Matrix (F-term):}
	Having obtained the $(k-b) \times (a-b)$ matrix $Q_t$ for the
	original variety $D$, we proceed with ease. Reading from the
	extraneous column of FI-parametres, we recognise matrices $Q$ 
	(corresponding to the rows that have zero in the extraneous
	column) and $V \cdot U$ (corresponding to those with
	combinations of the unresolved $\zeta$'s in the last column).
	We let $V \cdot U$ be $c \times (a-b)$ whereby
	making $Q$ of dimension $(k-b-c) \times (a-b)$. The number $c$
 	is easily read from the embedding of $D$ into $D'$ as the
	number of unresolved FI-parametres.\\
	From $Q$, we compute the kernel $T$, a matrix of dimensions 
	$(a-b)-(k-b-c) \times (a-b) = (a-k+c) \times (a-b)$ as well as
	the matrix $K^t$ of dimensions $(a-k+c) \times d$ describing the
	dual cone to that spanned by the columns of $T$. The integer
	$d$ is uniquely determined from the dimensions of $T$ in
	accordance with the algorithm of finding dual cones presented
	in the Appendix. From these two matrices we compute $T^t \cdot
	K^t$, of dimension $(a-b) \times d$.
\item \underline{Obtaining the $\Delta$ Matrix (D-term):}
	Finally, we use (\ref{Delta}) to compute
	$(V \cdot U) \cdot (T^t \cdot K^t)$, arriving
	at our desired matrix $\Delta$ of dimensions $c \times d$,
	the incidence matrix of our quiver diagram. The number of
	gauge groups we have is therefore $c+1$ and the
	number of bi-fundamentals, $d$.\\
	Of course one may dispute that finding the kernel $T$ of $Q$
	is highly non-unique as
	any basis change in the null-space would give an equally valid $T$.
	This is indeed so. However we note that it is really the
 	combination $T^t \cdot K^t$ that we need. This is a dot-product
	in disguise, and by the very definition of
	the dual cone, this combination remains invariant under basis changes.
	Therefore this step of obtaining the quiver $\Delta$ from the
	charge matrix $Q_t$ is a unique	procedure.
\end{enumerate}
\subsection{Obtaining the Superpotential}
Having noticed that the matter content can be conveniently obtained,
we proceed to address the interactions, i.e., the
F-terms, which require a little more care. The matrix $K$ which our
algorithm extracts encodes the F-term equations and must at least be such
that they could be integrated back to a single function: the
{\it superpotential.}

Reading the possible F-flatness equations from $K$ is {\it ipso facto}
straight-forward. The subtlety exists in how to find the right
candidate among many different linear relations.
As mentioned earlier, $K$ has dimensions $m \times (r-2)$ with $m$
corresponding to the fields that will finally manifest in the
superpotential, $r-2$, the fields that solve them according to
(\ref{kmatrix}) and (\ref{p_i}); of course, $m \ge r-2$.
Therefore we have
$r-2$ vectors in $\Z^m$, giving generically $m-r+2$ linear relations
among them. Say we have ${\rm row}_1 + {\rm row}_3 -  {\rm row}_7 =
0$, then we simply write down $X_1 X_3 = X_7$ as one of the candidate
F-terms.
In general, a relation $\sum\limits_i a_i K_{ij} = 0$ with $a_i \in
\Z$ implies an F-term
$\prod\limits_i X_i^{a_i} = 1$ in accordance with (\ref{kmatrix}).
Of course, to find all the linear relations, we simply find the
$\Z$-nullspace of $K^t$ of dimension $m-r+2$.

Here a great ambiguity exists, as in our previous calculations of
nullspaces: any linear combinations therewithin may suffice to give a
new relation as a candidate F-term\footnote{Indeed each linear
	relation gives a possible
	candidate and we seek the correct ones. For the sake of
	clarity we shall call candidates ``relations'' and reserve the
	term ``F-term'' for a successful candidate.}.
Thus educated guesses are called for in order to find
the set of linear relations which may be most conveniently integrated
back into the superpotential. Ideally, we wish this back-integration
procedure to involve no extraneous fields (i.e., integration
constants\footnote{By constants we really mean functions since we are
	dealing with systems of partial differential equations.})
other than the $m$ fields which appear in the K-matrix. Indeed, as we shall
see, this wish may not always be granted and sometimes 
we must include new fields. In this case,
the whole moduli space of the gauge theory will be larger than
the one encoded by our toric data and the new fields parametrise new
branches of the moduli in the theory.

Let us return to the SPP example to enlighten ourselves before generalising.
We recall from subsection 3.2, that ${\scriptsize K=\left(
	\begin{array}{c|cccccc}
		& X_1 & X_2 & X_3 & X_4 & X_5 & X_6 \\ \hline
		v_1 & 1 & 0 & 0 & 0 & 0 & 0 \cr
		v_2 & 0 & 0 & 1 & 0 & 1 & 0 \cr
		v_3 & 0 & 1 & 0 & 0 & 0 & 0 \cr
		v_4 & 0 & 0 & 1 & 1 & 0 & 0 \cr
		v_5 & 0 & 0 & 0 & 1 & 0 & 1  \end{array}
\right)}$ from which we can read out only one relation
$X_3 X_6 - X_4 X_5=0$ using the rule described in the paragraph
above. Of course there can be only one relation because the nullspace
of $K^t$ is of dimension $6-5=1$.

Next we must calculate the charge under the gauge groups which this
term carries. We must ensure that the superpotential, being a term in a
Lagrangian, be a gauge invariant, i.e., carries no overall charge
under $\Delta$.
From ${\scriptsize d = \left(
	\begin{array}{c|cccccc}
	& X_1 &  X_2 & X_3 & X_4 & X_5 & X_6 \\ \hline
U(1)_A  & -1 &   1 &  0 & 1 &  -1 &  0 \\
U(1)_B  & 1  &  -1 &  1 & 0 &   0 & -1 \\
U(1)_C  & 0  &   0 &  -1& -1 &  1 &  1 \end{array} \right)}$
we find the charge of $X_3 X_6$ to be $(q_A,q_B,q_C) =
(0+0,1+(-1),(-1)+1) = (0,0,0)$; of course by our very construction,
$X_4 X_5$ has the same charge. Now we have two choices:
(a) to try to write the superpotential using only the six fields; or 
(b) to include some new field $\phi$ which also has charge $(0,0,0)$.
For (a) we can try the ansatz  $W=X_1 X_2 (X_3 X_6 -X_4  X_5)$ which
does give our F-term upon partial derivative with respect to $X_1$ or
$X_2$. However, we would also have a new F-term $X_1 X_2 X_3=0$ by
${\partial \over {\partial X_6}}$, which is inconsistent with our $K$
since columns 1, 2 and 3 certainly do not add to 0.

This leaves us with option (b), i.e., $W= \phi (X_3 X_6 -X_4  X_5)$ say.
In this case, when $\phi=0$ we not only obtain our F-term,
we need not even correct the matter content $\Delta$. This branch of
the moduli space is that of our original theory.
However, when $\phi \neq 0$, we must have $X_3=X_4=X_5=X_6=0$. 
Now the D-terms read $|X_1|^2 - |X_2|^2 = -\zeta_1 = \zeta_2$,
so the moduli space is: $\{\phi \in \C, X_1 \in \C \}$ such that
$\zeta_1+\zeta_2=0$ for otherwise there would be no moduli at all.
We see that we obtain another branch of moduli space.
As remarked before, this is a general phenomenon when we include
new fields: the whole moduli space will be larger than the one encoded
by the toric data. As a check, we see that our example is exactly that
given in \cite{Uranga}, after the identification with their notation,
$Y_{12} \rightarrow X_6, X_{24}\rightarrow X_3, Z_{23}\rightarrow X_1,
Z_{32}\rightarrow X_2, Y_{34}\rightarrow X_4, X_{13}\rightarrow X_5,
Z_{41}\rightarrow\phi$ and $(X_1 X_2 - \phi) \rightarrow \phi$.
We note that if we were studying a non-Abelian extension to the toric
theory, as by brane setups (e.g. \cite{Uranga}) or by stacks
of probes (in progress from \cite{Chris}),
the chargeless field $\phi$ would manifest
as an adjoint field thereby modifying our quiver diagram. Of course since
the study of toric methods in physics is so far restricted to product
$U(1)$ gauge groups, such complexities do not arise. To avoid
confusion we shall henceforth mark only the bi-fundamentals in our quiver
diagrams but will write the chargeless fields explicit in the
superpotential.

Our agreement with the results of \cite{Uranga} is very reassuring. It
gives an excellent example demonstrating that our canonical resolution
technique and the inverse algorithm do indeed, in response to what was
posited earlier, give a theory living on
a D-brane probing the SPP (T-dual to the setup in \cite{Uranga}).
However, there is a subtle point we would like to mention. There
exists an ambiguity in writing the superpotential when the chargeless
field $\phi$ is involved. Our algorithm gives $W = \phi (X_3 X_6 - X_4
X_6)$ while \cite{Uranga} gives $W = (X_1 X_2 - \phi)(X_3 X_6 - X_4
X_6)$. Even though they have identical moduli, it is the latter which
is used for the brane setup. Indeed, the toric methods by definition 
(in defining $\Delta$ from $a_{ij}$) do not handle chargeless fields
and hence we have ambiguities. Fortunately our later
examples will not involve such fields.

The above example of the SPP was a na\"{\i}ve one as we need only to
accommodate a single F-term. We move on to a more complicated example.
Suppose we are now given
${\tiny
d = \left(
\matrix{ & X_1&X_2&X_3&X_4&X_5&X_6&X_7&X_8&X_9&X_{10} \cr
 A&-1&0&0&-1&0&0&0&1&0&1 \cr B&1&-1&0&0&0&-1&0&0&1&0 \cr C&0&0&1&
  0&1&0&1&-1&-1&-1 \cr D&0&1&-1&1&-1&1&-1&0&0&0 \cr  }
\right)
}$ and
${\tiny
K = \left(
\matrix{
X_1&X_2&X_3&X_4&X_5&X_6&X_7&X_8&X_9&X_{10} \cr
 1&0&1&0&0&0&1&0&0&0 \cr 0&1&1&0&0&0&0&1&0&0 \cr 
1&0&0&1&0&0&0&0&1&0 \cr 0&1&0&1&0&1&0&0
  &0&0 \cr 0&0&1&0&1&0&1&0&0&0 \cr 0&0&0&0&0&1&1&0&0&1 \cr  }
\right)
}$. We shall see in the next section, that these arise for
the del Pezzo 1 surface. 
Now the nullspace of $K$ has dimension $10-6=4$, we could obtain a host of
relations from various linear combinations in this space.
One relation is obvious: $X_2 X_7 - X_3 X_6 = 0$. The charge it
carries is $(q_A,q_B,q_C,q_D) = (0+0,-1+0,0+1,1+(-1)) = (0,-1,1,0)$
which cancels that of $X_9$. Hence $X_9 (X_2 X_7 - X_3 X_6)$ could be
a term in $W$. Now ${\partial \over{\partial X_2}}$ thereof gives $X_7
X_9$ and from $K$ we see that $X_7 X_9 - X_1 X_5 X_{10} = 0$,
therefore, $- X_1 X_2 X_5 X_{10}$ could be another term in $W$. We repeat
this procedure, generating new terms as we proceed and introducing new
fields where necessary. We are fortunate that in this case we can
actually reproduce all F-terms without recourse to artificial
insertions of new fields: $W = X_{2} X_{7} X_{9} - X_{3} X_{6} X_{9}
- X_{4} X_{8} X_{7} - X_{1} X_{2} X_{5} X_{10} + X_{3} X_{4} X_{10} 
+ X_{1} X_{5} X_{6} X_{8}$.

Enlightened by these examples, let us return to some remarks upon
generalities.
Making all the exponents of the fields positive, the F-terms can then
be written as 
\beq
\label{F-term}
\prod\limits_i X_i^{a_i} = \prod\limits_j X_j^{b_j},
\eeq
with $a_i, b_j \in \Z^+$. Indeed if we were to have another field
$X_k$ such that $k \not\in \{i\}, \{j\}$ then the term
$X_k \left(\prod\limits_i X_i^{a_i} -  \prod\limits_j X_j^{b_j}
\right)$, on the condition that $X_k$ appears only this once, must be an
additive term in the superpotential $W$. This is because the
F-flatness condition ${\partial W \over {\partial {X_k}}} = 0$ implies
(\ref{F-term}) immediately. Of course judicious observations are called
for to (A) find appropriate relations (\ref{F-term}) and (B) find 
$X_k$ among our $m$ fields. Indeed (B) may not even be possible and
new fields may be forced to be introduced, whereby making the moduli
space of the gauge theory larger than that encodable by the toric data.

In addition, we must ensure that each term in $W$ be chargeless under
the product gauge groups.
What this means for us is that 
for each of the terms $X_k \left(\prod\limits_i X_i^{a_i} -  
\prod\limits_j X_j^{b_j}\right)$ we must have
${\rm Charge}_s(X_k) + \sum\limits_i a_i {\rm Charge}_s(X_i) = 0$
for $s=1,..,r$ indexing through our $r$ gauge group factors (we note
that by our very construction, for each gauge group, the charges 
for $\prod\limits_i X_i^{a_i}$ and for $\prod\limits_j X_j^{b_j}$ 
are equal).
If $X_k$ in fact cannot be found among our $m$ fields, it must be
introduced as a new field $\phi$ with appropriate charge.
Therefore with each such relation (\ref{F-term}) read from $K$, we
iteratively perform this said procedure, checking
$\Delta_{sk} + \sum\limits_i a_i \Delta_{si} = 0$ at each step, until
a satisfactory superpotential is reached. The right choices throughout
demands constant vigilance and astuteness.

\section{An Illustrative Example: the Toric del Pezzo Surfaces}
As the $\C^3 / (\Z_2 \times \Z_2)$ resolutions were studied in great
detail in \cite{Uranga}, we shall use the data from \cite{Chris} to
demonstrate the algorithm of finding the gauge theory from toric
diagrams extensively presented in the previous section.

\begin{figure}
\centerline{\psfig{figure=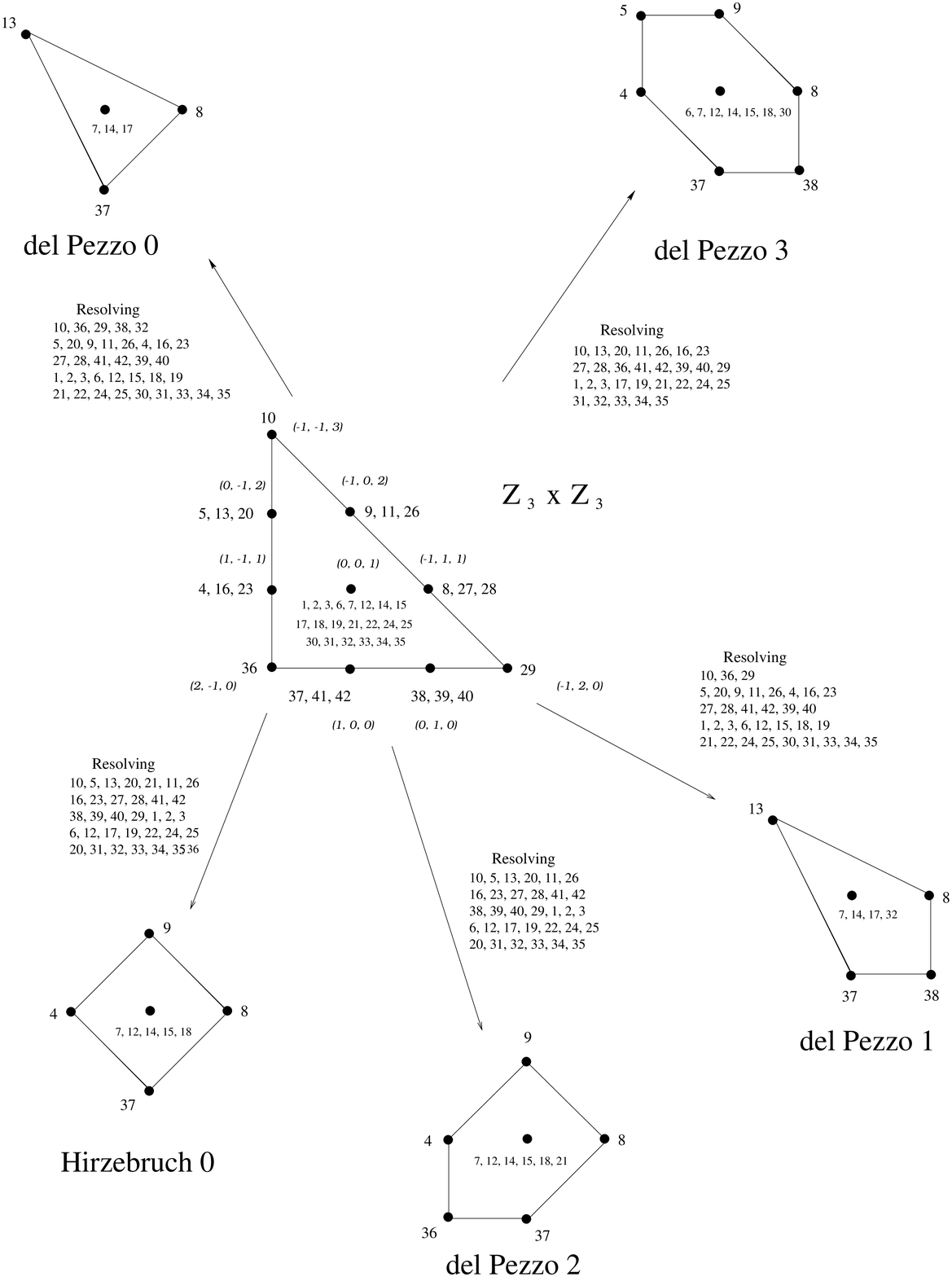,width=5.0in}}
\caption{The resolution of the Gorenstein singularity $\C^3/(\Z_3
	\times \Z_3)$ to the three toric del Pezzo surfaces as well as
	the zeroth Hirzebruch surface. We have
	labelled explicitly which columns (linear $\sigma$-model
	fields) are to be associated to each node in the toric
	diagrams and especially which columns are to be eliminated
	(fields acquiring non-zero VEV) in the various
	resolutions. Also, we have labelled the nodes of the parent
	toric diagram with the coordinates as given in the matrix
	$G_t$ for $\C^3/(\Z_3 \times \Z_3)$.}
\label{f:dP}
\end{figure}

The toric diagram of the dual cone of the (parent) quotient
singularity $\C^3 / (\Z_3 \times \Z_3)$ as well as those of its
resolution to the three toric del Pezzo surface are presented in
\fref{f:dP}.

\begin{figure}
\centerline{\psfig{figure=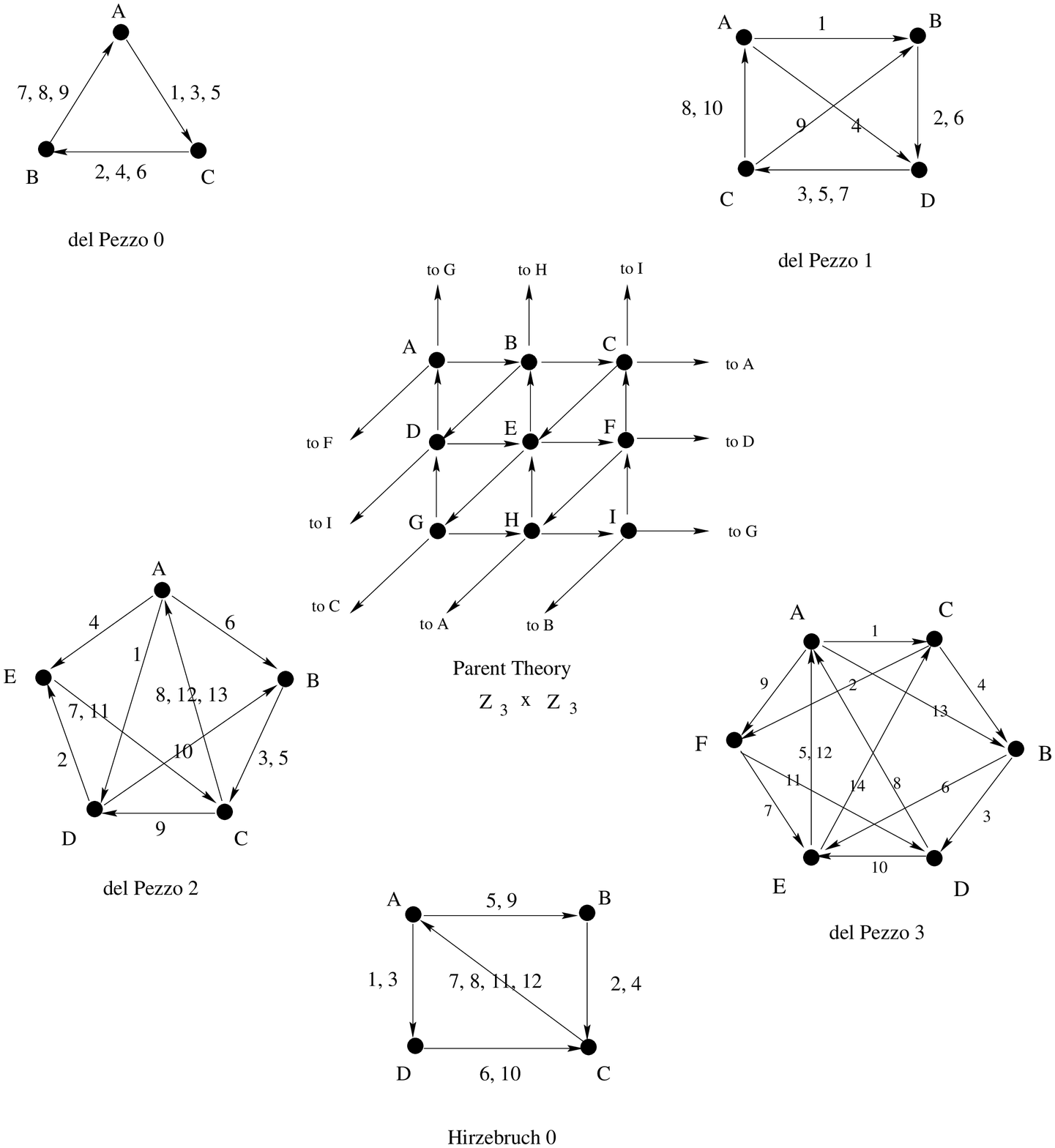,width=6.0in}}
\caption{The quiver diagrams for the matter content of the
	brane world-volume gauge theory on the 4 toric del Pezzo
	singularities as well as the zeroth Hirzebruch surface.
	We have specifically labelled the $U(1)$ 
	gauge groups (A, B, ..) and the bi-fundamentals (1, 2, ..)
	in accordance with our conventions in presenting the various
	matrices $Q_t$, $\Delta$ and $K$. As a reference we have also
	included the quiver for the parent $\Z_3 \times \Z_3$ theory.}
\label{f:dPquiver}
\end{figure}

\underline{del Pezzo 1:}
Let us commence our analysis with the first toric del Pezzo
surface\footnote{Now some may identify the toric diagram of del Pezzo
	1 as given by nodes (using the notation in \fref{f:dP}) $(1,-1,1)$, 
	$(2,-1,0)$, $(-1,1,1)$, $(0,0,1)$ and $(-1,0,2)$ instead of
	the one we have chosen in the convention of \cite{Chris},
	with nodes  $(0,-1,2)$, $(0,0,1)$, $(-1,1,1)$, $(1,0,0)$ and
	$(0,1,0)$. But of course these two $G_t$ matrices describe the
	same algebraic variety. The former corresponds to
	${\rm Spec}\left(\C[XY^{-1}Z,X^2Y^{-1},X^{-1}YZ,Z,X^{-1}Z^2]\right)$ 
	while the latter corresponds to
	${\rm Spec}\left(\C[Y^{-1}Z^2,Z,X^{-1}YZ,X,Y]\right)$. The
	observation that $(X^2Y^{-1})=(X)(X^{-1}YZ)^{-1}(Z)$,
	$(XY^{-1}Z) = (X)(Y)^{-1}(Z)$ and $(X^{-1}Z^2) =
	(Y^{-1}Z^2)(Y)(X^{-1})$ for the generators of the polynomial
	ring gives the equivalence. In other words, there is an
	$SL(5,\Z)$ transformation between the 5 nodes of the two toric
	diagrams.}. From its toric diagram, we see that the minimal
$\Z_n \times\Z_n$ toric diagram into which it embeds is $n=3$. As a
reference, the toric diagram for $\C^3 / (\Z_3 \times \Z_3)$ is given
in \fref{f:dP} and the quiver diagram, given later in the convenient
brane-box form, in \fref{f:dPquiver}. Luckily, the
matrices $Q'_t$ and $G'_t$ for this Abelian quotient is given in
\cite{Chris}. Adding the extra column of FI-parametres we present
these matrices below\footnote{In \cite{Chris}, a canonical
	ordering was used; for our purposes we need not belabour this point
	and use their $Q'_{total}$ as $Q'_t$. This is perfectly legitimate as
	long as we label the columns carefully, which we have done.}:
\[
{\tiny
\begin{array}{l}
G'_t = \left(
\matrix{p_{1}& p_{2}& p_{3}& p_{4}& p_{5}& p_{6}& p_{7}&
	p_{8}& p_{9}& p_{10}& p_{11}& p_{12}& p_{13}& 
  	p_{14}& p_{15}& p_{16}& p_{17}& p_{18}& p_{19}& p_{20} 
	& p_{21}& p_{22}& p_{23}& p_{24} \cr 
0&0&0&1&0&0&0&-1&-1&-1&-1&0&0&0&0&1&0&0&0&0&0&0&1&0
   \cr 0&0&0&-1&-1&0&0&1&0&-1&0&0&-1&0&0&-1&0&0&0&-1&0&0&-1&0
   \cr 1&1&1&1&2&1&1&1&2&3&2&1&2&1&1&1&1&1&1&2&1&1&1&1 \cr  }
\right. \cdots \cdots
\\ \\ \\
\qquad \qquad \qquad \qquad \cdots \cdots
\qquad \qquad
\left.
\matrix{
 p_{25}& p_{26}& p_{27}& p_{28}& p_{29}& p_{30}& p_{31}& p_{32}& 
p_{33}& p_{34}& p_{35}& p_{36}&  p_{37}& p_{38}& p_{39}& p_{40}
& p_{41}& p_{42} & \cr
 0&-1&-1&-1&-1&0&0&0&0&0&0&2&1&0&0&0&1&1 \cr 0&0&1&1&2&0
  &0&0&0&0&0&-1&0&1&1&1&0&0 \cr 1&2&1&1&0&1&1&1&1&1&1&0&0&
  0&0&0&0&0 \cr  }
\right)
\end{array}
}
\]
and
\[
{\tiny
\begin{array}{l}
Q'_t =
\left(
\matrix{  p_{1}& p_{2}& p_{3}& p_{4}& p_{5}& p_{6}& p_{7}&
	p_{8}& p_{9}& p_{10}& p_{11}& p_{12}& p_{13}& 
  	p_{14}& p_{15}& p_{16}& p_{17}& p_{18}& p_{19}& p_{20} 
	& p_{21}& p_{22}& p_{23}& p_{24}& \cr 
	1&0&0&0&0&0&0&0&0&0&0&0&0&-1&0&0&0&0&0&0&1&0&0&0&  0  \cr
  0&1&0&0&0&0&0&0&0&0&0&0&0&-1&0&0&0&0&0&0&1&0&0&0&  0  \cr0&0
  &1&0&0&0&0&0&0&0&0&0&0&-1&0&0&0&0&0&0&1&0&0&0&  0  \cr0&0&0&
  1&0&0&0&0&0&0&0&0&0&-1&0&0&0&0&0&0&1&0&0&0&  0  \cr0&0&0&0&1
  &0&0&0&0&0&0&0&0&-1&0&0&0&0&0&0&1&0&0&0&  0  \cr0&0&0&0&0&1&
  0&0&0&0&0&0&0&-1&0&0&0&0&0&0&1&0&0&0&  0  \cr0&0&0&0&0&0&1&0
  &0&0&0&0&0&-1&0&0&0&0&0&0&1&0&0&0&  0  \cr0&0&0&0&0&0&0&1&0&
  0&0&0&0&-1&0&0&0&0&0&0&0&0&0&0&  0  \cr0&0&0&0&0&0&0&0&1&0&0
  &0&0&-1&0&0&0&0&0&0&0&0&0&0&  0  \cr0&0&0&0&0&0&0&0&0&1&0&0&
  0&-1&0&0&0&0&0&0&0&0&0&0&  0  \cr0&0&0&0&0&0&0&0&0&0&1&0&0&-1
  &0&0&0&0&0&0&0&0&0&0&  0  \cr0&0&0&0&0&0&0&0&0&0&0&1&0&-1&0&
  0&0&0&0&0&0&0&0&0&  0  \cr0&0&0&0&0&0&0&0&0&0&0&0&1&-1&0&0&0
  &0&0&0&0&0&0&0&  0  \cr0&0&0&0&0&0&0&0&0&0&0&0&0&0&1&0&0&0&0
  &0&-1&0&0&0&  0  \cr0&0&0&0&0&0&0&0&0&0&0&0&0&0&0&1&0&0&0&0&
  -1&0&0&0&  0  \cr0&0&0&0&0&0&0&0&0&0&0&0&0&0&0&0&1&0&0&0&-1&0
  &0&0&  0  \cr0&0&0&0&0&0&0&0&0&0&0&0&0&0&0&0&0&1&0&0&-1&0&0&
  0&  0  \cr0&0&0&0&0&0&0&0&0&0&0&0&0&0&0&0&0&0&1&0&-1&0&0&0&  0  \cr0
  &0&0&0&0&0&0&0&0&0&0&0&0&0&0&0&0&0&0&1&-1&0&0&0&  0  \cr0&0&
  0&0&0&0&0&0&0&0&0&0&0&0&0&0&0&0&0&0&0&1&0&0&  0  \cr0&0&0&0&
  0&0&0&0&0&0&0&0&0&0&0&0&0&0&0&0&0&0&1&0&  0  \cr0&0&0&0&0&0&
  0&0&0&0&0&0&0&0&0&0&0&0&0&0&0&0&0&1&  0  \cr0&0&0&0&0&0&0&0&
  0&0&0&0&0&0&0&0&0&0&0&0&0&0&0&0&  0  \cr0&0&0&0&0&0&0&0&0&0&
  0&0&0&0&0&0&0&0&0&0&0&0&0&0&  0  \cr0&0&0&0&0&0&0&0&0&0&0&0&
  0&0&0&0&0&0&0&0&0&0&0&0&  0  \cr0&0&0&0&0&0&0&0&0&0&0&0&0&0&
  0&0&0&0&0&0&0&0&0&0&  0  \cr0&0&0&0&0&0&0&0&0&0&0&0&0&0&0&0&
  0&0&0&0&0&0&0&0&  0  \cr0&0&0&0&0&0&0&0&0&0&0&0&0&0&0&0&0&0&
  0&0&0&0&0&0&  0  \cr0&0&0&0&0&0&0&0&0&0&0&0&0&0&0&0&0&0&0&0&
  0&0&0&0&  0  \cr0&0&0&0&0&0&0&0&0&0&0&0&0&0&0&0&0&0&0&0&0&0&
  0&0&  0  \cr0&0&0&0&0&0&0&0&0&0&0&0&0&0&0&0&0&0&0&0&0&0&0&0
  &  0  \cr-1&1&0&0&0&0&0&0&0&0&0&0&0&0&0&0&0&0&0&0&0&0&0&0& \zeta_{1} \cr0
  &1&-1&-1&2&-1&0&1&0&-1&0&0&0&0&-1&1&0&0&0&0&0&0&0&0& \zeta_{2} \cr0&
  1&0&0&0&-1&0&0&0&0&0&0&0&0&0&0&0&0&0&0&0&0&0&0& \zeta_{3} \cr1&-2&1
  &0&-2&2&0&-1&0&1&0&0&0&0&0&0&0&0&0&0&0&0&0&0& \zeta_{4} \cr-1&1&-1&1
  &0&-1&0&1&0&0&0&0&0&0&0&0&0&0&0&0&0&0&0&0& \zeta_{5} \cr0&0&0&1&-2
  &1&0&-1&0&1&0&0&0&0&0&-1&0&0&0&0&0&1&0&0& \zeta_{6} \cr0&-1&1&0&0&0
  &0&0&0&0&0&0&0&0&0&0&0&0&0&0&0&0&0&0& \zeta_{7} \cr0&1&-1&0&2&-1&0
  &1&0&-1&0&0&0&0&0&0&0&0&0&0&0&-1&0&0& \zeta_{8} \cr }
\right. \cdots \cdots
\\ \\ \\
\qquad \qquad \qquad \qquad \cdots \cdots
\qquad \qquad
\left.
\matrix{
 p_{25}& p_{26}& p_{27}& p_{28}& p_{29}& p_{30}& p_{31}& p_{32}& 
p_{33}& p_{34}& p_{35}& p_{36}&  p_{37}& p_{38}& p_{39}& p_{40}
& p_{41}& p_{42} & \cr 
 -1&0&0&0&0&0&0&0&0&1&-1&0&0&0&1&-1&-1&1&  0  \cr-1&0&0&0&0&
  0&0&0&1&0&-1&0&0&0&1&-1&-1&1&  0  \cr-1&0&0&0&0&0&0&-1&2&0&-1&0
  &0&0&2&-2&-2&2&  0  \cr-1&0&0&0&0&0&0&-1&1&1&-1&0&0&0&2&-1&-2&1
  &  0  \cr-2&0&0&0&0&0&0&-1&2&1&-2&0&0&0&3&-2&-2&2&  0  \cr-2&0&0&0&0&0
  &0&0&2&0&-1&0&0&0&1&-1&-1&1&  0  \cr-2&0&0&0&0&0&0&0&1&1&-1&0&
  0&0&1&-1&-1&1&  0  \cr-1&0&0&0&0&0&0&0&1&0&0&0&0&0&0&-1&0&1&  0  \cr-1
  &0&0&0&0&0&0&0&0&1&-1&0&0&0&1&-1&0&1&  0  \cr-1&0&0&0&0&0&0&-1
  &1&0&-1&0&0&0&2&-1&-1&2&  0  \cr-1&0&0&0&0&0&0&-1&2&-1&0&0&0&0
  &1&-1&-1&2&  0  \cr0&0&0&0&0&0&0&-1&1&0&0&0&0&0&1&-1&-1&1&  0  \cr0&0
  &0&0&0&0&0&-1&0&0&0&0&0&0&1&0&-1&1&  0  \cr1&0&0&0&0&0&0&0&-1
  &0&0&0&0&0&-1&1&1&-1&  0  \cr1&0&0&0&0&0&0&0&-1&-1&1&0&0&0&-1&
  2&0&-1&  0  \cr1&0&0&0&0&0&0&-1&0&-1&1&0&0&0&-1&1&0&0&  0  \cr1&0&0&0
  &0&0&0&-1&0&0&0&0&0&0&0&0&0&0&  0  \cr0&0&0&0&0&0&0&0&0&-1&1
  &0&0&0&-1&1&0&0&  0  \cr0&0&0&0&0&0&0&0&0&-1&0&0&0&0&0&1&0&0
  &  0  \cr-1&0&0&0&0&0&0&0&0&1&-1&0&0&0&1&-1&0&0&  0  \cr-1&0&0&0&0&0
  &0&0&1&0&-1&0&0&0&1&0&-1&0&  0  \cr-1&0&0&0&0&0&0&0&1&-1&0&0&0
  &0&0&0&0&0&  0  \cr0&1&0&0&0&0&0&-1&0&0&-1&0&0&0&1&-1&0&1&  0  \cr0&
  0&1&0&0&0&0&-1&1&-1&0&0&0&0&0&-1&0&1&  0  \cr0&0&0&1&0&0&0&0&
  -1&1&-1&0&0&0&0&-1&1&0&  0  \cr0&0&0&0&1&0&0&0&0&0&0&0&0&0&-1&
  -1&1&0&  0  \cr0&0&0&0&0&1&0&-1&0&1&-1&0&0&0&1&-1&0&0&  0  \cr0&0&0&0
  &0&0&1&-1&1&0&-1&0&0&0&1&-1&-1&1&  0  \cr0&0&0&0&0&0&0&0&0&0&0
  &1&0&0&0&1&-1&-1&  0  \cr0&0&0&0&0&0&0&0&0&0&0&0&1&0&-1&1&0&-1
  &  0  \cr0&0&0&0&0&0&0&0&0&0&0&0&0&1&-1&0&1&-1&  0  \cr0&0&0&0&0&0&0
  &0&0&0&0&0&0&0&0&0&0&0& \zeta_{1} \cr0&0&0&0&0&0&0&0&0&0&0&0&0&0&0
  &0&0&0& \zeta_{2} \cr0&0&0&0&0&0&0&0&0&0&0&0&0&0&0&0&0&0& \zeta_{3} \cr0&0&0&0
  &0&0&0&0&0&0&0&0&0&0&0&0&0&0& \zeta_{4} \cr0&0&0&0&0&0&0&0&0&0&0&0
  &0&0&0&0&0&0& \zeta_{5} \cr0&0&0&0&0&0&0&0&0&0&0&0&0&0&0&0&0&0& \zeta_{6} \cr0
  &0&0&0&0&0&0&0&0&0&0&0&0&0&0&0&0&0& \zeta_{7} \cr0&0&0&0&0&0&0&0&0
  &0&0&0&0&0&0&0&0&0&\zeta_{8} \cr }
\right)
\end{array}
}
\]

According to our algorithm, we must perform Gaussian row-reduction on
$Q'_t$ to solve for 42 variables $x_i$. When this is done we find that we
can in fact express all variables in terms of 3 $x_i$'s together with
the 8 FI-parametres $\zeta_i$. We choose these three $x_i$'s to be
$x_{10,29,36}$ corresponding to the 3 outer vertices which we know must be
resolved in going from $\C^3 / (\Z_3 \times \Z_3)$ to del Pezzo 1.

Next we select the fields which must be kept and set them to zero in
order to determine the range for $\zeta_i$. Bearing in mind the toric
diagrams from \fref{f:dP}, these fields we
judiciously select to be: $p_{13, 8, 37, 38}$.
Setting $x_{13,8,37,38} = 0$  gives us the solution
$\{ \zeta_6=0; x_{29} = \zeta_7 = \zeta_3 = \zeta_1 - \zeta_5;
x_{10} = \zeta_4 + \zeta_5 + \zeta_3; x_{36} = \zeta_7 - \zeta_8
\}$, which upon back-substitution to the solutions $x_i$ we obtained
from $Q'_t$, gives zero for $x_{13, 8, 37, 38}$ (which we have
chosen by construction) as well as $x_{7,14,17,32}$; 
for all others we obtain positive
values. This means precisely that all the other fields
are to be eliminated and these 8 columns 
\{ 13, 8, 37, 38, 7,14,17,32 \}
are to be kept while the remaining 42-8=34 are to
be eliminated from $Q'_t$ upon row-reduction to give $Q_t$.
In other words, we have found our set $B$ to be
\{1,2,3,4,5,6,9,10,11,12,15,16,18,19,20,21,22,23,24,25,
26,27,28,29,30,31,33,34,35,36,39,40,41,42\}
and thus according to (\ref{qt}) we immediately obtain
\[
{\scriptsize
Q_t = \left(
\matrix{ 
p_{7}& p_{8}& p_{13}& p_{14}& p_{17}& p_{32}& p_{37}& p_{38} & \cr
1&0&0&0&0&-1&0&0&\zeta_2 + \zeta_8 \cr 0&0&0&0&-1&1&0&0&\zeta_6 \cr -1&0&
  0&1&0&0&0&0&\zeta_1 + \zeta_3 + \zeta_5 \cr 0&0&1&-1&0&-1&0&1&0 \cr -1&1&1&-1&-1
  &0&1&0&0 \cr  }
\right).
}
\]
We note of course that 5 out of the 8 FI-parametres have been
eliminated automatically; this is to be expected since in resolving 
$\C^3 / (\Z_3 \times \Z_3)$ to del Pezzo 1, we remove precisely 5
nodes. Obtaining the D-terms and F-terms is now straight-forward.
Using (\ref{Delta}) and re-inserting the last row we obtain the
D-term equations (incidence matrix) to be
\[
{\scriptsize
d = \left(
\matrix{ X_1&X_2&X_3&X_4&X_5&X_6&X_7&X_8&X_9&X_{10} \cr
 -1&0&0&-1&0&0&0&1&0&1 \cr 1&-1&0&0&0&-1&0&0&1&0 \cr 0&0&1&
  0&1&0&1&-1&-1&-1 \cr 0&1&-1&1&-1&1&-1&0&0&0 \cr  }
\right)
}
\]
From this matrix we immediately observe that there are 4 gauge groups,
i.e., $U(1)^4$ with 10 matter fields $X_i$ which we have labelled in
the matrix above. In an equivalent notation we rewrite $d$ as the
adjacency matrix of the quiver diagram (see \fref{f:dPquiver})
for the gauge theory:
\[
{\scriptsize
a_{ij} = \left(
\matrix{0&0&2&0\cr 1&0&1&0\cr 0&0&0&3\cr 1&2&0&0}
\right).
}
\]
The K-matrix we obtain to be:
\[
{\scriptsize
K^t = \left(
\matrix{
X_1&X_2&X_3&X_4&X_5&X_6&X_7&X_8&X_9&X_{10} \cr
 1&0&1&0&0&0&1&0&0&0 \cr 0&1&1&0&0&0&0&1&0&0 \cr 
1&0&0&1&0&0&0&0&1&0 \cr 0&1&0&1&0&1&0&0
  &0&0 \cr 0&0&1&0&1&0&1&0&0&0 \cr 0&0&0&0&0&1&1&0&0&1 \cr  }
\right)
}
\]
which indicates that the original 10 fields $X_i$ can be expressed in
terms of 6. This was actually addressed in the previous section and we
rewrite that pleasant superpotential here:
\[
W = X_{2} X_{7} X_{9} - X_{3} X_{6} X_{9} - X_{4} X_{8} 
	X_{7} - X_{1} X_{2} X_{5} X_{10} + X_{3} X_{4} X_{10} 
	+ X_{1} X_{5} X_{6} X_{8}.
\]

\underline{del Pezzo 2:}
Having obtained the gauge theory for del Pezzo 1, we now repeat the
above analysis for del Pezzo 2. Now we have the
FI-parametres restricted as
$\{ p_{36} = \zeta_2 = 0; \zeta_3 = \zeta_4; x_{29} = \zeta_4
+\zeta_6; x_{10} = \zeta_1 + \zeta_4 \}$, making the set to be
eliminated as $B = \{$ 1, 2, 3, 5, 6, 10, 11, 13, 16, 17, 19, 20,
22, 23, 24, 25, 26, 27, 28, 29, 30, 31, 32, 33, 34, 35, 
38, 39, 40, 41, 42 $\}$. Whence, we obtain
\[
{\scriptsize
Q_t = \left(
\matrix{
p_{4} & p_{7} & p_{8} & p_{9} & p_{12} & p_{14} & p_{15}
 & p_{18} & p_{21} & p_{36} & p_{37} & \cr
 0&1&0&0&0&0&0&0&-1&0&0&\zeta_{4} + \zeta_{6} + \zeta_{8} \cr 
1&-1&1&0&0&-1&0&0& 0&0&0&\zeta_{7} \cr
 0&-1&0&0&0&1&0&0&0&0&0&\zeta_{1} + \zeta_{3} + \zeta_{5} \cr -1&1&-1&0&0
  &1&-1&0&1&0&0&\zeta_{2} \cr 0&-1&0&1&0&0&-1&0&0&0&1&0 \cr 0&-1&1&1&0
  &-1&0&0&-1&1&0&0 \cr -1&1&-1&0&0&1&-1&1&0&0&0&0 \cr -1&1&-1&0&1
  &0&0&0&0&0&0&0 \cr  }
\right),
}
\]
and observe that 4 D-terms have been resolved, as 4 nodes have been
eliminated from $\C^3 / (\Z_3 \times \Z_3)$.
From this we easily extract (see \fref{f:dPquiver})
\[
{\scriptsize
d = 
\left(
\matrix{
X_{1} & X_{2} & X_{3} & X_{4} & X_{5} & X_{6} & X_{7} & X_{8} & X_{9} & 
   X_{10} & X_{11} & X_{12} & X_{13} \cr
 -1 & 0 & 0 & -1 & 0 & -1 & 0 & 1 & 0 & 0 & 0 & 1 & 1 \cr 0 & 0 & -1
    & 0 & -1 & 1 & 0 & 0 & 0 & 1 & 0 & 0 & 0 \cr 0 & 0 & 1 & 0 & 1 & 0 & 1 & 
   -1 & -1 & 0 & 1 & -1 & -1 \cr 1 & -1 & 0 & 0 & 0 & 0 & 0 & 0 & 1 & -1 & 0
    & 0 & 0 \cr 0 & 1 & 0 & 1 & 0 & 0 & -1 & 0 & 0 & 0 & -1 & 0 & 0 \cr  } 
\right);
}
\]
moreover, we integrate the F-term matrices
\[
{\scriptsize
K^t =
\left(
\matrix{
X_{1} & X_{2} & X_{3} & X_{4} & X_{5} & X_{6} & X_{7} & X_{8} & X_{9} & 
   X_{10} & X_{11} & X_{12} & X_{13} \cr
 0&1&1&0&0&0&0&1&0&0&0&1&0 \cr 1&0&1&0&0&0&0&0&0&0&1&1
  &0 \cr 1&0&0&1&0&1&0&0&1&0&0&0&0 \cr 0&1&0&1&0&1&0&0&0&1&0&0
  &0 \cr 0&1&1&1&1&0&0&0&0&0&0&0&0 \cr 0&0&1&0&1&0&1&0&0&0&1&0
  &0 \cr 0&0&0&1&1&0&0&0&1&0&0&0&1 \cr  }
\right)
}
\]
to obtain the superpotential
\[
\begin{array}{c}
W = X_{2} X_{9} X_{11} - X_{9} X_{3} X_{10} - X_{4} X_{8} X_{11} -
X_{1} X_{2} X_{7} X_{13} + X_{13} X_{3} X_{6} \\
- X_{5} X_{12} X_{6}+
X_{1} X_{5} X_{8} X_{10} + X_{4} X_{7} X_{12}.
\end{array}
\]

\underline{del Pezzo 3:}
Finally, we shall proceed to treat del Pezzo 3.
Here we have the range of the FI-parametres to be
$\{ \zeta_1 = \zeta_6 = \zeta_6 = 0; x_{29} = \zeta_3 = -\zeta_5;
x_{10} = \zeta_4; \zeta_2 = x_{36}; \zeta_8 = -\zeta_2 - \zeta_{10}
\}$, which gives the set $B$ as
\{1, 2, 3, 10, 11, 13, 16, 17, 19, 20, 21, 22, 23, 24, 25, 
26, 27, 28, 29, 31, 32, 33, 34, 35, 36, 39, 40, 41, 42\}, 
and thus according to (\ref{qt}) we immediately obtain
\[
{\scriptsize
Q_t = \left(
\matrix{
p_{4}&p_{5}&p_{6}&p_{7}&p_{8}&p_{9}&p_{12}&p_{14}&p_{15}&p_{18}
&p_{30}&p_{37}&p_{38} &\cr
 0&0&0&1&0&0&0&0&0&0&-1&0&0&\zeta_2  + \zeta_4  + \zeta_8  \cr 1&0&0&-1&1&0
  &0&-1&0&0&0&0&0&\zeta_7  \cr -1&0&0&1&-1&0&0&1&-1&0&1&0&0&\zeta_6  \cr 0
  &0&-1&0&0&0&0&1&0&0&0&0&0&\zeta_3  + \zeta_5  \cr 0&0&1&-1&0&0&0&0&0&0
  &0&0&0&\zeta_1  \cr 0&1&-1&0&0&0&0&0&0&0&-1&0&1&0 \cr -1&1&-1&0&0&0
  &0&1&-1&0&0&1&0&0 \cr -1&0&0&1&-1&0&0&1&-1&1&0&0&0&0 \cr -1&0
  &0&1&-1&0&1&0&0&0&0&0&0&0 \cr 1&-1&1&-1&0&1&0&-1&0&0&0&0&0&0 \cr  }
\right)
}
\]
We note indeed that 3 out of the 8 FI-parametres have been
automatically resolved, as we have removed 3 nodes from the toric
diagram for $\C^3 / (\Z_3 \times \Z_3)$.
The matter content (see \fref{f:dPquiver}) is encoded in
\[
{\scriptsize
d = \left(
\matrix{
 X_{1} & X_{2} & X_{3} & X_{4} & X_{5} & X_{6} & X_{7} & X_{8} & X_{9} & 
   X_{10} & X_{11} & X_{12} & X_{13} & X_{14} \cr 
 -1&0&0&0&1&0&0&1&-1&0&0&1&-1&0 \cr 0&0
   &-1&1&0&-1&0&0&0&0&0&0&1&0 \cr 1&-1&0&-1&0
   &0&0&0&0&0&0&0&0&1 \cr 0&0&1&0&0&0&0&-1&0
   &-1&1&0&0&0 \cr 0&0&0&0&-1&1&1&0&0&1&0&-1
   &0&-1 \cr 0&1&0&0&0&0&-1&0&1&0&-1&0&0&0 \cr 
    }
\right),
}
\]
and from the F-terms
\[
{\scriptsize
K^t = \left(
\matrix{
 X_{1} & X_{2} & X_{3} & X_{4} & X_{5} & X_{6} & X_{7} & X_{8} & X_{9} & 
   X_{10} & X_{11} & X_{12} & X_{13} & X_{14} \cr 
 1&0&0&0&0&1&1&1&0&0&0&0&0&0 \cr 0&1&0&0&0&1&0&1&0&0&0&
  1&0&0 \cr 1&0&0&0&0&0&0&0&1&0&0&0&1&1 \cr 0&1&0&1&0&0&0&0&1&0
  &0&0&1&0 \cr 0&1&1&0&0&1&0&0&1&0&0&0&0&0 \cr 0&0&1&0&0&1&1&0
  &0&0&1&0&0&0 \cr 0&0&1&0&1&0&0&0&1&0&0&0&0&1 \cr 0&0&0&0&0&1
  &1&1&0&1&0&0&0&0 \cr  }
\right)
}
\]
we integrate to obtain the superpotential
\[
\begin{array}{c}
W = X_{3} X_{8} X_{13} - X_{8} X_{9} X_{11} - X_{5} X_{6} X_{13} - 
X_{1} X_{3} X_{4} X_{10} X_{12} \\
+ X_{7} X_{9} X_{12} + X_{1} X_{2} X_{5} X_{10} X_{11} + 
X_{4} X_{6} X_{14} - X_{2} X_{7} X_{14}.
\end{array}
\]
Note that we have a quintic term in $W$; this is an interesting
interaction indeed.

\underline{del Pezzo 0:}
Before proceeding further, let us attempt one more example, viz., the
degenerate case of the del Pezzo 0 as shown in \fref{f:dP}. This time
we note that the ranges for the FI-parametres are $\{
\zeta_{5}=-x_{29}+\zeta_{6}-A; \zeta_{6}=x_{29}-B; x_{29}=B+C; 
\zeta_{8}=-x_{36}+B; x_{36}=B+C+D; x_{10}=A+E \}$ for some positive $A,
B, C, D$ and $E$, that $B =$ \{
1, 2, 3, 4, 5, 6, 9, 10, 11, 12, 15, 16, 18, 19, 20, 21, 22, 
23, 24, 25, 26, 27, 28, 29, 30, 31, 32, 33, 34, 35, 36, 38, 
39, 40, 41, 42 \} and whence the charge matrix is
\[
{\scriptsize
Q_t = \left(
\matrix{
p_{7} & p_{8} & p_{13} & p_{14} & p_{17} & p_{37} & \cr
 1&0&0&0&-1&0&\zeta_{2} + \zeta_{6} + \zeta_{8} \cr
-1&0&0&1&0&0&\zeta_{1} + \zeta_{3} + \zeta_{5} \cr -1&1&1&-1&-1&1&0
\cr  }
\right).
}
\]
We extract the matter content (see \fref{f:dPquiver}) as
$
{\scriptsize
d = \left(
\matrix{
X_{1}& X_{2}& X_{3}& X_{4}& X_{5}& X_{6}& X_{7}& X_{8}& X_{9} \cr
-1 & 0 & -1 & 0 & -1 & 0 & 1 & 1 & 1 \cr
0 & 1 & 0 & 1 & 0 & 1 & -1 & -1 & -1 \cr
1 & -1 & 1 & -1 & 1 & -1 & 0 & 0 & 0 \cr}
\right),
}
$
and the F-terms as
$
{\scriptsize
K^t = \left(
\matrix{ 
X_{1}& X_{2}& X_{3}& X_{4}& X_{5}& X_{6}& X_{7}& X_{8}& X_{9} \cr
1 & 1 & 0 & 0 & 0 & 0 & 1 & 0 & 0 \cr
1 & 0 & 1 & 0 & 1 & 0 & 0 & 0 & 0 \cr
 0 & 1 & 0 & 1 & 0 & 1 & 0 & 0 & 0 \cr 
0 & 0 & 1 & 1 & 0 & 0 & 0 & 1 & 0 \cr 0 & 
  0 & 0 & 0 & 1 & 1 & 0 & 0 & 1 \cr  }
\right),
}
$
and from the latter we integrate to obtain the superpotential
\[
W = X_{1} X_{4} X_{9} - X_{4} X_{5} X_{7} - X_{2} X_{3} X_{9} - 
	X_{1} X_{6} X_{8} + X_{2} X_{5} X_{8} + X_{3} X_{6} X_{7}.
\]
Of course we immediately recognise the matter content (which gives a
triangular quiver which we shall summarise below in \fref{f:dPquiver})
as well as the
superpotential from equations (4.7-4.14) of \cite{DGM}; it is simply
the theory on the Abelian orbifold $\C^3/\Z_3$ with action 
$(\alpha \in \Z_3) : (z_1, z_2, z_3) \rightarrow (e^{{2 \pi i} \over
3} z_1, e^{{2 \pi i} \over 3} z_2, e^{{2 \pi i} \over 3} z_3)$. 
Is our del Pezzo 0
then $\C^3/\Z_3$? We could easily check from the $G_t$ matrix (which
we recall is obtained from $G'_t$ of $\C^3 / (\Z_3 \times \Z_3)$
by eliminating the columns corresponding to the set $B$):
\[
{\scriptsize
G_t = \left(
\matrix{ 0 & -1 & 0 & 0 & 0 & 1 \cr 0 & 1 & -1 & 0 & 0 & 0 \cr
1 & 1 & 2 & 1 & 1 & 0 \cr  }
\right).
}
\]
These columns (up to repeat) correspond to monomials $Z, X^{-1}YZ,
Y^{-1}Z^2,X$ in the polynomial ring $\C[X,Y,Z]$. Therefore we need to find the
spectrum (set of maximal ideals) of the ring $\C[Z,
X^{-1}YZ,Y^{-1}Z^2,X]$, which is given by the minimal polynomial
relation: $(X^{-1}YZ) \cdot (Y^{-1}Z^2)\cdot X = (Z)^3$. This means,
upon defining $p = X^{-1}YZ; q = Y^{-1}Z^2; r=X$ and $s=Z$, our del
Pezzo 0 is described by $p q r = s^3$ as an algebraic variety 
in $\C^4({p,q,r,s})$, which is precisely $\C^3/\Z_3$. Therefore we
have actually come through a full circle in resolving $\C^3 / (\Z_3
\times \Z_3)$ to $\C^3/\Z_3$ and the validity of our algorithm
survives this consistency check beautifully. Moreover, since we know
that our gauge theory is exactly the one which lives on a D-brane probe on
$\C^3/\Z_3$, this gives a good check for physicality: that our careful
tuning of FI-parametres via canonical partial resolutions does give a
physical D-brane theory at the end.
We tabulate the matter content $a_{ij}$ and the superpotential $W$ for
the del Pezzo surfaces below, and the quiver diagrams, in \fref{f:dPquiver}.
\[
{\scriptsize
\begin{array}{|c|c|c|c|}
\hline
& \mbox{{\bf del Pezzo 1}} & \mbox{{\bf del Pezzo 2}} & \mbox{{\bf del Pezzo 3}} \\
\hline
\mbox{Matter } a_{ij}= &
	\left(\matrix{0&0&2&0\cr 1&0&1&0\cr 0&0&0&3\cr
		1&2&0&0}\right)&
	\left(\matrix{ 0 & 1 & 0 & 1 & 1 \cr 0 & 0 & 2 & 0 & 0 \cr
		 3 & 0 & 0 & 1 & 0 \cr 0
    		& 1 & 0 & 0 & 1 \cr 0 & 0 & 2 & 0 & 0 \cr} \right) &
	\left(\matrix{ 0&0&1&1&0&1 \cr 0&0&0&1&1&0\cr
	 	0 & 1 & 0 & 0& 0 & 1 \cr 1 & 0 & 0 & 0 & 1 & 0 \cr
		2 & 0 & 1 & 0 & 0 & 0 \cr 0 & 0 & 0
    		& 1 & 1 & 0 \cr  } \right)
\\ \hline
\mbox{Superpotential } W=  & \begin{array}{c}
	X_{2} X_{7} X_{9} - X_{3} X_{6} X_{9} \\
	- X_{4} X_{8} X_{7} - X_{1} X_{2} X_{5} X_{10} \\
	+ X_{3} X_{4} X_{10} + X_{1} X_{5} X_{6} X_{8} \end{array} &
	\begin{array}{c}
	X_{2} X_{9} X_{11} - X_{9} X_{3} X_{10} \\
	- X_{4} X_{8} X_{11} - X_{1} X_{2} X_{7} X_{13} \\
	+ X_{13} X_{3} X_{6} - X_{5} X_{12} X_{6} \\
	+ X_{1} X_{5} X_{8} X_{10} + X_{4} X_{7} X_{12}
	\end{array} &
	\begin{array}{c}
	X_{3} X_{8} X_{13} - X_{8} X_{9} X_{11} \\
	- X_{5} X_{6} X_{13} - X_{1} X_{3} X_{4} X_{10} X_{12} \\
	+ X_{7} X_{9} X_{12} + X_{1} X_{2} X_{5} X_{10} X_{11} \\
	+ X_{4} X_{6} X_{14} - X_{2} X_{7} X_{14}
	\end{array}
\\
\hline
\end{array}
}
\]
\[
{\scriptsize
\begin{array}{|c|c|c|}
\hline
& \mbox{{\bf del Pezzo 0}} \cong \C^3/\Z_3 & 
  \mbox{{\bf Hirzebruch 0}} \cong \P^1 \times \P^1 := F_0 = E_1 \\
\hline
\mbox{Matter }a_{ij} & \left(\matrix{0&3&0 \cr 0&0&3 \cr 3&0&0\cr}\right) &
	\left(\matrix{ 0 & 2 & 0 & 2 \cr 0 & 0 & 2 & 0 \cr 
	4 & 0 & 0 & 0 \cr 0 & 0 & 2 & 0 \cr  }\right)
\\ \hline
\mbox{Superpotential }W & \begin{array}{c}
	X_{1} X_{4} X_{9} - X_{4} X_{5} X_{7} \\
	- X_{2} X_{3} X_{9} - X_{1} X_{6} X_{8} \\
	+ X_{2} X_{5} X_{8} + X_{3} X_{6} X_{7}
	\end{array} &
	\begin{array}{c}
	 X_{1}X_{8}X_{10}- X_{3}X_{7}X_{10} \\
	- X_{2}X_{8}X_{9}- X_{1}X_{6}X_{12} \\
	+ X_{3}X_{6}X_{11}+ X_{4}X_{7}X_{9} \\
	+ X_{2}X_{5}X_{12}- X_{4}X_{5}X_{11}
	\end{array}
\\ \hline
\end{array}
}
\]
Upon comparing \fref{f:dP} and \fref{f:dPquiver}, we notice that as we go
from del Pezzo 0 to 3, the number of points in the toric diagram
increases from 4 to 7, and the number of gauge groups (nodes in the
quiver) increases from 3 to 6. This is consistent with the observation
for ${\cal N} = 1$ theories that the number of gauge groups equals the
number of perimetre points (e.g., for del Pezzo 1, the four nodes 13,
8, 37 and 38) in the toric diagram. Moreover, as discussed in
\cite{AHK}, the rank of the global symmetry group ($E_i$ for del Pezzo
$i$) which must exist for
these theories equals the number of perimetre point minus 3; it would
be an intereting check indeed to see how such a symmetry manifests
itself in the quivers and superpotentials.

\underline{Hirzebruch 0:}
Let us indulge ourselves with one more example, namely the 0th
Hirzebruch surface, or simply $\P^1 \times \P^1 := F_0 := E_1$.
The toric diagram is drawn in \fref{f:dP}. Now the FI-parametres are
$\{ \zeta_4 = -x_{29}-x_{36}-\zeta_5-\zeta_8-A; \zeta_5 =
-A-B; \zeta_7 = x_{10}+x_{29}+x_{36}+\zeta_8-C; \zeta_8 =
-x_{10}-x_{29}-x_{36}+D; D = A+B; C = A+B; A = x_{10}-E; x_{10} = E+F;
x_{29} = B+G \}$ for positive $A,B,C,D,E,F$ and $G$. Moreover, $B=$ \{
1, 2, 3, 5, 6, 10, 11, 13, 16, 17, 19, 20, 21, 22, 23, 24, 25, 26, 27,
28, 29, 30, 31, 32, 33, 34, 35, 36, 38, 39, 40, 41, 42 \}. We note
that this can be obtained directly by partial resolution of fields 
21 and 36 from
del Pezzo 2 as is consistent with \fref{f:dP}. Therefrom we obtain the
charge matrix
\[
{\scriptsize
Q_t = \left(
\matrix{
p_{4}& p_{7}& p_{8}& p_{9}& p_{12}& p_{14}& p_{15}& p_{18}& p_{37} & \cr
 -1&2&-1&0&0&1&-1&0&0&\zeta_2 + \zeta_4 + \zeta_6 + \zeta_8 \cr
1&-1&1&0&0&-1&0&0&0&\zeta_7 \cr
 0&-1&0&0&0&1 &0&0&0&\zeta_1 + \zeta_3 + \zeta_5 \cr 
0&-1&0&1&0&0&-1&0&1&0 \cr 
-1&1&-1&0&0&1&-1&1&0&0 \cr
-1&1&-1&0&1&0&0&0&0&0 \cr  }
\right),
}
\]
from which we have the matter content
${\scriptsize
d = \left(
\matrix{
X_{1}& X_{2}& X_{3}& X_{4}& X_{5}& X_{6}& X_{7}& X_{8}& X_{9}& 
X_{10}& X_{11}& X_{12} \cr
-1 & 0 & -1 & 0 & -1 & 0 & 1 & 1 & -1 & 0 & 1 & 1 \cr 
0 & -1 & 0 & -1 & 1 & 0 & 0 & 0 & 1 & 0 & 0 & 0 \cr 
0 & 1 & 0 & 1 & 0 & 1 & -1 & -1 & 0 & 1 & -1 & -1 \cr  
1 & 0 & 1 & 0 & 0 & -1 & 0 & 0 & 0 & -1 & 0 & 0 \cr}
\right)
}$ the quiver for which is presented in \fref{f:dPquiver}.
The F-terms are
\[{\scriptsize
K^t = \left(
\matrix{
X_{1}& X_{2}& X_{3}& X_{4}& X_{5}& X_{6}& X_{7}& X_{8}& X_{9}& 
X_{10}& X_{11}& X_{12} \cr
1 & 1 & 0 & 0 & 0 & 0 & 1 & 0 & 0 & 0 & 1 & 0 \cr 
1 & 0 & 1 & 0 & 1 & 0 & 0 & 0 & 1 & 0 & 0 & 0 \cr 
1 & 1 & 1 & 1 & 0 & 0 & 0 & 0 & 0 & 0 & 0 & 0 \cr
0 & 1 & 0 & 1 & 0 & 1 & 0 & 0 & 0 & 1 & 0 & 0 \cr
0 & 0 & 1 & 1 & 0 & 0 & 0 & 1 & 0 & 0 & 0 & 1 \cr
0 & 0 & 0 & 0 & 1 & 1 & 1 & 1 & 0 & 0 & 0 & 0 \cr  } \right)
},\]
from which we obtain
\[
W = X_{1}X_{8}X_{10}- X_{3}X_{7}X_{10}- X_{2}X_{8}X_{9}- X_{1}X_{6}X_{12}+ 
X_{3}X_{6}X_{11}+ X_{4}X_{7}X_{9}+ X_{2}X_{5}X_{12}- X_{4}X_{5}X_{11},
\]
a perfectly acceptable superpotential with only cubic interactions. We include
these results with our table above.

\section{Uniqueness?}
In our foregoing discussion we have constructed in detail an algorithm
which calculates the matter content encoded by $\Delta$ and superpotential
encoded in $K$, given the toric diagram of the singularity which the
D-branes probe. As abovementioned, though this algorithm gives
one solution for the quiver and the $K$-matrix once the matrix $Q_t$
is determined, the general inverse
process of going from toric data to gauge theory data, 
is highly {\bf non-unique} and a classification of all possible
theories having the same toric description would be
interesting\footnote{We thank R. Plesser for pointing this issue out
	to us.}.
Indeed, by the very structure of our
algorithm, in immediately appealing to the partial resolution of gauge
theories on $\Z_n \times \Z_n$ orbifolds which are well-studied, we
have granted ourselves enough extraneous information to determine a
unique $Q_t$ and hence the ability to proceed with ease (this was the
very reason for our devising the algorithm).

However, generically we do not have any such luxury.
At the end of subsection 3.1, we have already mentioned two types of
ambiguities in the inverse problem. Let us refresh our minds. They
were (A) the {\bf F-D ambiguity} which is the inability to decide, simply by
observing the toric diagram, which rows of the charge matrix $Q_t$ are
D-terms and which are F-terms and (B) the {\bf repetition ambiguity} which
is the inability to decide which columns of $G_t$ to repeat once
having read the vectors from the toric diagram. Other ambiguities
exist, such as in each time when we compute nullspaces, but we shall here
discuss to how ambiguities (A) and (B) manifest themselves and
provide examples of vastly different gauge theories having the same
toric description. There is another point which we wish to emphasise: 
as mentioned at the end of subsection
3.1, the resolution method guarantees, upon careful tuning of the
FI-parametres, that the resulting gauge theory does originate from the
world-volume of a D-brane probe. Now of course, by taking liberties
with experimentation of these ambiguities we are no longer protected by
physicality and in general the theories no longer live on the D-brane.
It would be a truly interesting exercise to check which
of these different theories do.

\underline{F-D Ambiguity:}
First, we demonstrate type (A) by returning to our old
friend the SPP whose charge matrix we had earlier presented.
Now we write the
same matrix without specifying the FI-parametres:
\[
{\scriptsize
Q_t = \left(
\matrix{ 1 & -1 & 1 & 0 & -1 & 0 \cr -1 & 1 & 0 & 1 & 0 & -1 \cr
	 -1 & 0 & 0 & -1 & 1 & 1 \cr  } 
\right)
}
\]
We could apply the last steps of our algorithm to this matrix as
follows.
\begin{enumerate}
\def\theenumi{\alph{enumi}}\def\labelenumi{(\theenumi)}
\item If we treat the first row as $Q$ (the F-terms) and the second and
	third as $V \cdot U$ (the D-terms) we obtain the gauge theory
	as discussed in subsection 3.3 and in \cite{Uranga}.
\item If we treat the second row as $Q$ and first with the third as
	$V \cdot U$, we obtain ${\scriptsize d = \left(
	\matrix{ -1 & 0 & 1 & -1 & 1 & 0 \cr 
		1 & 0 & 0 & 1 & -2 & -1 \cr
		0 & 0 & -1 & 0 & 1 & 1}
	\right)}$ which is an exotic theory indeed with a field ($p_5$)
	charged under three gauge groups.\\
	Let us digress a moment to address the stringency of the
	requirements upon matter contents. By the
	very nature of finite group representations, any orbifold theory
	must give rise to only adjoints and bi-fundamentals because its matter
	content is encodable by an adjacency matrix due to tensors of
	representations of finite groups.
	The corresponding incidence matrix $d$, has (a) only 0 and $\pm 1$
	entries specifying the particular bi-fundamentals and (b) has
	each column containing precisely one 1, one $-1$ and
	with the remaining entries 0. However more exotic
	matter contents could arise from more generic toric
	singularities, such as fields charged under 3
	or more gauge group factors; these would then have $d$ matrices with
	conditions (a) and (b) relaxed\footnote{Note that we still
		require that each column sums to 0 so as to be able to	
		factor out an overall $U(1)$.}.
	Such exotic quivers (if we could even
	call them quivers still) would give
	interesting enrichment to those well-classified families as discussed
	in \cite{Quiver}.\\
	Moreover we must check the anomaly cancellation
	conditions. These could be rather involved; even though for
	$U(1)$ theories they are a little simpler, we still need to
	check {\em trace anomalies} and {\em cubic anomalies}.
	In a trace-anomaly-free theory, for each node in the
	quiver, the number of incoming arrows must equal the number of
	outgoing (this is true for a $U(1)$ theory which is what
	toric varieties provide; for a discussion on this see
	e.g. \cite{Han-He}). In matrix language this means that each
	row of $d$ must sum to 0.\\
	Now for a theory with only bi-fundamental matter with $\pm 1$
	charges, since $(\pm 1)^3 = \pm 1$, the cubic is equal
	to the trace anamaly; therefore for these theories we need
	only check the above row-condition for $d$. For more exotic
	matter content, which we shall meet later, we do need to
	perform an independent cubic-anomaly check.\\
	Now for the above $d$, the second row does not sum to zero
	and whence we do unfortunately have a
	problematic anomalous theory. Let us push on to see whether
	we have better luck in the following.
\item Treating row 3 as the F-terms and the other two as the D-terms
	gives\\
	${\scriptsize d = \left(
	\matrix{ 0 & -1 & 1 & -1 & 1 & 0 \cr
		0 & 1 & 0 & 1 & -2 & -1 \cr 
		0 & 0 & -1 & 0 & 1 & 1}
	\right)}$ which has the same anomaly problem as the one above.
\item Now let rows 1 and 2 as the F-terms and the 3rd, as the D-terms,
	we obtain ${\scriptsize d = \left(
	\matrix{X_1 & X_2 & X_3 & X_4 & X_5 \cr
		0 & 1 & 1 & -1 & -1 \cr
		0 & -1 & -1 & 1 & 1}
	\right)}$, which is a perfectly reasonable matter content.
	Integrating
	${\scriptsize K = \left(
	\matrix{ 1 & 0 & 1 & 0 & 0 \cr 0 & 1 & 0 & 1 & 0 \cr
	1 & 0 & 0 & 1 & 0 \cr 0 & 0 & 1 & 0 & 1 \cr  }
	\right)}$ gives the
	superpotential $W = \phi(X_1 X_2 X_5 - X_3 X_4)$ for some
	field $\phi$ of charge $(0,0)$ (which could be an adjoint for
	example; note however that we can not use $X_1$ even though it
	has charge $(0,0)$ for otherwise the F-terms would be
	altered). This theory is perfectly legitimate. We compare the
	quiver diagrams of theories (a) (which we recall from
	\fref{f:SPPquiver}) and this present example in
	\fref{f:compare}. As a check, let us define the gauge
	invariant quantities: $a=X_2 X_4$, $b=X_2 X_5$, $c=X_3 X_4$,
	$d=X_3 X_5$ and $e=X_1$. Then we have the algebraic relations 
	$ad=bc$ and $eb = c$, from which we immediately obtain $ad =
	eb^2$, precisely the equation for the SPP.
\begin{figure}
\centerline{\psfig{figure=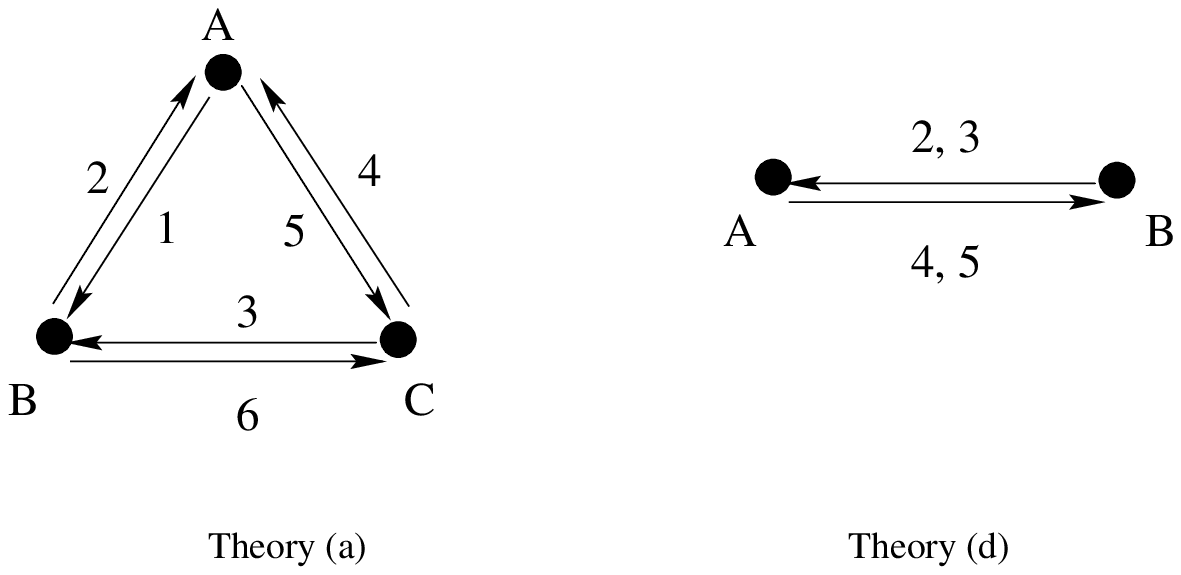,width=5.0in}}
\caption{The vastly different matter contents of theories (a) and (d),
	both anomaly free and flow to the toric diagram of the
	suspended pinched point in the IR.}
\label{f:compare}
\end{figure}
\item As a permutation on the above, treating rows 1 and 3 as the
	F-terms gives a theory equivalent thereto.
\item Furthermore, we could let rows 2 and 3 be $Q$ giving us
	${\scriptsize d = \left(
	\matrix{0 & 1 & -1 & -1 & -1 \cr
		0 & -1 & 1 & 1 & 1}
	\right)}$, but this again gives an anomalous matter content.
\item Finally, though we cannot treat all rows as F-terms, we can
	however treat all of them as D-terms in which $Q_t$ is simply
	$\Delta$ as remarked at the end of Section 2 before the flow
	chart. In this case we have the matter content
	${\scriptsize d = \left(
	\matrix{ 1 & -1 & 1 & 0 & -1 & 0 \cr -1 & 1 & 0 & 1 & 0 & -1 \cr
	 -1 & 0 & 0 & -1 & 1 & 1 \cr  1 & 0 & -1 & 0 & 0 & 0} 
	\right)}$ which clearly is both trace-anomaly free (each row
	adds to zero) and cubic-anomaly-free (the cube-sum of the each
	row is also zero). The superpotential, by our very choice, is
	of course zero. Thus we have a perfectly legitimate theory
	without superpotential but with an exotic field (the first column)
	charged under 4 gauge groups.			
\end{enumerate}
We see therefore, from our list of examples above, that for the simple
case of the SPP we have 3 rather different theories (a,d,g) with
contrasting matter content and superpotential which
share the same toric description.

\underline{Repetition Ambiguity:}
As a further illustration, let us give one example of type (B)
ambiguity.
First let us eliminate all repetitive columns from the $G_t$ of SPP,
giving us:
\[
{\scriptsize G_t = \left(
\matrix{ 1 & 0 & 0 & -1 & 1 \cr 1 & 1 & 0 & 1 & 0 \cr 
1 & 1 & 1 & 1 & 1 \cr} \right), }
\]
which is perfectly allowed and consistent with \fref{f:SPP}.
Of course many more possibilities for
repeats are allowed and we could redo the following analyses for each
of them.
As the nullspace of our present choice of $G_t$, 
we find $Q_t$, and we choose, in light
of the foregoing discussion, the first row to represent the D-term:
\[
{\scriptsize Q_t = \left(
\matrix{ -1 & 1 & -1 & 0 & 1 & \zeta \cr 1 & -2 & 0 & 1 & 0 & 0 \cr}
\right).}
\]
Thus equipped, we immediately retrieve, using our algorithm,
\[
{\scriptsize
d = \left(
	\matrix{X_1 & X_2 & X_3 & X_4 & X_5 \cr
		1 & -1 & 1 & -1 & 0 \cr -1 & 1 & -1 & 1 & 0}
\right)
\qquad
K^t = \left(
	\matrix{ 1 & 0 & 0 & 0 & 0\cr 0 & 0 & 2 & 0 & 1\cr
	0 & 1 & 0 & 0 & 0\cr 0 & 0 & 1 & 1 & 1\cr  }
\right)
\qquad
T = \left(
	\matrix{ 0 & 0 & 0 & 0 & 1 \cr -1 & 0 & 0 & 1 & 0 \cr
	0 & 0 & 1 & 0 & 0 \cr 2 & 1 & 0 & 0 & 0 \cr  } 
\right).
}
\]
We see that $d$ passes our anomaly test, with
the same bi-fundamental matter content as theory (d). The
superpotential can be read easily from $K$ (since there is only one
relation) as $W = \phi (X_5^2 - X_3 X_4)$.
As a check, let us define the gauge invariant quantities: $a=X_1 X_2$,
$b=X_1 X_4$, $c=X_3 X_2$, $d=X_3 X_4$ and $e=X_5$. These have among
themselves the
algebraic relations $ad=bc$ and $e^2 = d$, from which we immediately
obtain $bc = ae^2$, the equation for the SPP.
Hence we have yet another interesting anomaly free theory, 
which together with our theories (a), (d) and (g)
above, shares the toric description of the SPP.

Finally, let us indulge in one more demonstration. Now let us treat
both rows of our $Q_t$ as D-terms, whereby giving a theory with no
superpotential and the exotic matter content
${\scriptsize d = \left(
\matrix{ -1 & 1 & -1 & 0 & 1 \cr 1 & -2 & 0 & 1 & 0 \cr
	0 & 1 & 1 & -1 & -1}
\right)}$ with a field (column 2) charged under 3 gauge groups. Indeed
though the rows sum to 0 and trace-anomaly is avoided, the cube-sum of
the second row gives $1^3 + 1^3 + (-2)^3=-6$ and we do have a cubic
anomaly.

In summary, we have an interesting phenomenon indeed! Taking so immediate an
advantage of the ambiguities in the above has already produced quite a
few examples of vastly different gauge theories flowing in the IR to
the same universality class by having their moduli spaces identical.
The vigilant reader may raise two issues. First, as mentioned earlier,
one may take the pains to check whether these theories do indeed live on
a D-brane. Necessary conditions such as that the theories may be obtained from
an ${\cal N}=4$ theory must be satisfied. Second, the matching of
moduli spaces may not seem so strong since they are on a classical
level. However, since we are dealing with product $U(1)$ gauge groups
(which is what toric geometry is capable to dealing with so far), the
classical moduli receive no quantum corrections\footnote{We thank
	K. Intriligator for pointing this out.}. Therefore the
matching of the moduli for
these various theories do persist to the quantum regime, which hints
at some kind of ``duality'' in the field theory. We shall
call such a duality {\bf toric duality}. It would be interesting to
investigate how, with non-Abelian versions of the theory (either by
brane setups or stacks of D-brane probes), this toric duality may be
extended.

\section{Conclusions and Prospects}
The study of resolution of toric singularities by D-branes is by now
standard. In the concatenation of the F-terms and D-terms from the
world volume gauge theory of a single D-brane at the singularity, the
moduli space could be captured by the algebraic data of the toric
variety. However, unlike the orbifold theories, the inverse problem
where specifying the structure of the singularity specifies the
physical theory has not yet been addressed in detail.

We recognise that in contrast with D-brane probing orbifolds, where
knowing the group structure and its space-time action uniquely
dictates the matter content and superpotential, such flexibility is
not shared by generic toric varieties due to the highly non-unique
nature of the inverse problem. It has been the purpose and main
content of the current writing to device an {\bf algorithm} which
constructs the matter content (the incidence matrix $d$) and the
interaction (the F-term matrix $K$) of a well-behaved gauge theory 
given the toric diagram $D$ of the singularity at hand.

By embedding $D$ into the Abelian orbifold $\C^k/(\Z_n)^{k-1}$ and
performing the standard partial resolution techniques, we have
investigated how the induced action upon the charge matrices
corresponding to the toric data of the latter gives us a convenient
charge matrix for $D$ and have constructed a programmatic methodology
to extract the matter content and superpotential of {\it one} D-brane world
volume gauge theory probing $D$.
The theory we construct, having its origin from an orbifold, is nicely
behaved in that it is anomaly free, with bi-fundamentals only and
well-defined superpotentials.
As illustrations we have tabulated the results for
all the toric del Pezzo surfaces and the zeroth Hirzebruch surface.

Directions of further work are immediately clear to us. From the
patterns emerging from del Pezzo surfaces 0 to 3, we could speculate
the physics of higher (non-toric) del Pezzo cases. For example, we
expect del Pezzo $n$ to have $n+3$ gauge groups.
Moreover, we could attempt to fathom how our resolution techniques
translate as Higgsing in brane setups, perhaps with recourse to
diamonds, and realise the various theories
on toric varieties as brane configurations.

Indeed, as mentioned, the inverse problem is highly non-unique; we
could presumably attempt to classify all the different theories
sharing the same toric singularity as their moduli space. In light of
this, we have addressed two types of ambiguity: that in having
multiple fields assigned to the same node in the toric diagram and
that of distinguishing the F-terms and D-terms in the charge
matrix. In particular we have turned this ambiguity to a matter of
interest and have shown, using our algorithm, how vastly different theories,
some with quite exotic matter content, may have the same toric
description. This commonality would correspond to a duality wherein
different gauge theories flow to the same universality class in the
IR. We call this phenomenon {\bf toric duality}.
It would be interesting indeed how this duality may
manifest itself as motions of branes in the corresponding setups.
Without further ado however, let us pause here awhile and leave such
investigations to forthcoming work.

\section*{Appendix: Finding the Dual Cone}
Let us be given a convex polytope $C$,  with the edges specifying
the faces of which given by the matrix $M$ whose columns are the 
vectors corresponding to these edges.
Our task is to find the dual cone $\tilde{C}$ of $C$, or more
precisely the matrix $N$ such that
\[
N^t \cdot M \ge 0 \qquad \mbox{for all entries.}
\]
There is a standard algorithm, given in \cite{Fulton}.
Let $M$ be $n \times p$, i.e., there are $p$ $n$-dimensional vectors
spanning $C$. We note of course that $p \ge n$ for convexity.
Out of the $p$ vectors, we choose $n-1$. This gives us an $n \times
(n-1)$ matrix of co-rank 1, whence we can extract a 1-dimensional
null-space (as indeed the initial $p$ vectors are all linearly
independent) described by a single vector $u$.

Next we check the dot product of $u$ with the remaining $p-(n-1)$
vectors. If all the dot products are positive we keep $u$, and if
all are negative, we keep $-u$, otherwise we discard it.

We then select another $n-1$ vectors and repeat the above until all
combinations are exhausted. The set of vectors we have kept, $u$'s or
$-u$'s then form the columns of $N$ and span the dual cone
$\tilde{C}$.

We note that this is a very computationally intensive algorithm, the
number of steps of which depends on 
$\left( \begin{array}{c} p \\ n -1 \\ \end{array} \right)$ which
grows exponentially.

A subtle point to remark. In light of what we discussed in a footnote
in the paper on the difference between ${\bf M_+} = {\bf M} \cap
\sigma$ and ${\bf M'_+}$, here we have computed the dual of
$\sigma$. We must ensure that $\Z_+$-independent lattice points inside
the cones be not missed.

\section*{Acknowledgements}
We would like to extend our sincere gratitude to the CTP of MIT for
her gracious patronage as well as the Institute for Theoretical Phyics
at UCSB for her warm hospitality and for hosting the
``Program on Supersymmetric Gauge Dynamics and String Theory.''
Furthermore, we thank K. Intriligator and
J. S. Song for
insightful comments. AH is grateful to M. Aganagic,
D.-E. Diaconescu, A. Karch, D. Morrison and R. Plesser for
valuable discussions.
BF thanks A. Uranga and R. von Unge
for helpful insights. YHH acknowledges V. Rodoplu of Stanford
University and S. Wu of the Dept of Mathematics, UCSB for enlightening
discussions and is ever indebted to M. R. Warden for inspiration and
emotional support.


\end{document}